\documentclass[aps,prd,11pt,reprint,showpacs,superscriptaddress,floatfix]{revtex4-2}
\usepackage[T1]{fontenc}
\usepackage[utf8]{inputenc}
\usepackage{lmodern}
\usepackage{amsmath,amssymb,amsthm,physics}
\usepackage{graphicx,epstopdf}
\usepackage[dvipsnames]{xcolor}
\usepackage{float}
\usepackage{textgreek}
\usepackage{ulem}
\usepackage{orcidlink}
\usepackage{appendix}
\allowdisplaybreaks

\usepackage{xcolor}
\usepackage{booktabs}
\usepackage{graphicx}

\definecolor{navy}{RGB}{0,0,150}
\usepackage{appendix}
\allowdisplaybreaks
\newcommand{\RGU}{Department of Physics, The Assam Royal Global University, Guwahati-781035, Assam, India}
\newcommand{\ABTU}{Department of Physics, Al-Hussein Bin Talal University, 71111, Ma'an, Jordan}
\newcommand{\UCC}{Programa de P\'os-Gradua\c c\~ao em F\'{\i}sica \& Coordena\c c\~ao do Curso de F\'{\i}sica -- Bacharelado, Universidade Federal do Maranh\~{a}o, 65085-580 S\~{a}o Lu\'{\i}s, Maranh\~{a}o, Brazil}

\begin{document}

\title{ Screened Simpson-Visser Black Holes with Asymptotically de-Sitter Core }

\author{Faizuddin Ahmed\orcidlink{0000-0003-2196-9622}}
\email{faizuddinahmed15@gmail.com}
\affiliation{\RGU}

\author{Ahmad Al-Badawi\orcidlink{0000-0002-3127-3453}}
\email{ahmadbadawi@ahu.edu.jo}
\affiliation{\ABTU}

\author{Edilberto O. Silva\orcidlink{0000-0002-0297-5747}}
\email{edilberto.silva@ufma.br}
\affiliation{\UCC}

\begin{abstract}
In this work, we introduce a screened Simpson-Visser regular solution and perform a comprehensive study of its physical and observational properties. We begin by analyzing the thermodynamic behavior of the black hole, including detailed investigations of the Hawking temperature, Gibbs free energy, and specific heat, which provide insights into its stability and phase structure. Next, we examine the geodesic structure of the spacetime, considering both massless (photon) and massive (timelike) particles. In particular, we study the photon sphere, the corresponding black hole shadow, and the innermost stable circular orbits (ISCO), which are crucial for understanding the motion of matter and light around the black hole. Furthermore, we explore the black hole's energy-emission rate radiation, highlighting the effects of the modified geometry on observational signatures. Finally, we investigate the topological aspects of the black hole, analyzing both the thermodynamic topology and the photon sphere's topological properties. Our analysis demonstrates the intricate interplay between the spacetime geometry, geodesic motion, and black hole thermodynamics, offering a deeper understanding of this class of regular black holes and their potential observational consequences.
\end{abstract}

\maketitle


\section{Introduction}

One of the most striking predictions of classical general relativity is that sufficiently compact matter configurations inevitably undergo gravitational collapse to a spacetime singularity, a region where curvature invariants diverge, and geodesic completeness is lost.  This conclusion, enshrined in the singularity theorems of Penrose~\cite{Penrose1965} and Hawking~\cite{Hawking1970}, is widely regarded as a signal that GR breaks down at ultra-high energy densities and must be superseded by a quantum theory of gravity.  In the absence of such a complete theory, however, regular (singularity-free) black hole models offer a valuable phenomenological framework for probing the consequences of singularity resolution at the classical level.

The first regular black hole model was proposed by Bardeen~\cite{Bardeen1968}.  Although its field-theoretic interpretation remained elusive for decades, Ay\'on-Beato and Garc\'{\i}a~\cite{AyonBeato1998,AyonBeato2000} later showed that the Bardeen metric can be sourced by a nonlinear electromagnetic field satisfying the weak energy condition, giving it a solid physical foundation.  Shortly afterwards, they presented additional exact regular solutions within the same nonlinear electrodynamics (NLED) framework~\cite{AyonBeato1999}, and Bronnikov independently constructed a family of regular magnetic black holes and monopoles in NLED~\cite{Bronnikov2001}. Hayward's widely-cited model~\cite{Hayward2006} provided a dynamical description of regular black hole formation and evaporation, spurring a large literature on regular black holes sourced by de Sitter cores~\cite{Dymnikova1992,Fan2016}.  Reviews of these and related developments can be found in Refs.~\cite{Ansoldi2008,Lan2023}. Several regular black hole solutions, most of which were developed within the framework of general relativity, have been studied in \cite{Lobo2021, Frolov2016, Culetu2015a, Culetu2015b, Lemos2011, Dymnikova2004, Uchikata2012, Rodrigues2018, PonceDeLeon2017, Culetu2016, Balart2014}. In the context of $f(R)$ gravity, several approaches have been developed to study black holes. Specifically, regarding regular black holes, to the best of our knowledge, only a few works have addressed this topic \cite{Rodrigues2016a, Rodrigues2016b, Fabris2023}. 

A complementary class of regular spacetimes, the so-called \textit{black-bounce} geometries, was introduced by Simpson and Visser~\cite{Simpson2019}. The construction is remarkably simple: the areal radius $r$ in the Schwarzschild metric is replaced by $\sqrt{r^2+a^2}$, where the length scale $a$ regularizes the central singularity by replacing it with a smooth, minimum-area wormhole throat.  Depending on the ratio $a/(2M)$, the spacetime interpolates between a regular black hole (for $a < 2M$), a one-way wormhole (for $a = 2M$), and a two-way traversable wormhole (for $a > 2M$), all within a single, analytically tractable geometry.  Field sources for these spacetimes have been worked out in Refs.~\cite{Bronnikov2022,Rodrigues2022}, and the construction has been extended to rotating, charged, and higher-dimensional settings in a rapidly growing literature~\cite{Franzin2021,Lima2023a,Lima2023b,Alencar2024,Crispim2024}. The line-element describing the black-bounce geometries is given by
\begin{align}
ds^2&=-A(r,a)dt^2+\frac{dr^2}{A(r,a)}+D^2(r,a)\,d\Omega^2,\nonumber\\
A(r,a)&=1-\frac{2M}{\sqrt{r^2+a^2}},\qquad D(r,a)=\sqrt{r^2+a^2}.\label{visser}    
\end{align}
where $d\Omega^2=d\theta^2+\sin^2 \theta\,d\phi^2$ is the $2D$ unit solid sphere.\\

Independent of the specific choice of regularization scheme, another line of research has focused on modifying the \emph{fall-off} of the gravitational potential by exponentially screening the Newtonian mass term. Recently, Visser et al.~\cite{Alex2020} proposed a mass function of the form 
\(m(r) = m\, e^{-\eta/r},\) where $\eta$ is a parameter with dimensions of length. This leads to a lapse function \(A(r) = 1 - \frac{2 M}{r} e^{-\eta/r}, \) describing a regular spacetime with an asymptotically Minkowski core. It is worth noting that a closely related version of this model, where the parameter $\eta$ is associated with the electric charge, has previously been studied within the framework of nonlinear electrodynamics in both extremal and non-extremal cases \cite{Culetu2015a,Balart2014}. The exponential factor suppresses the curvature divergence at $r\to 0$ and introduces an effective screening length $\eta$ that interpolates between Schwarzschild (at $\eta=0$) and a geometry whose near-origin behaviour mimics that of Reissner-Nordstr\"om. The line-element describing the screened Schwarzschild geometry with asymptotically Minkowski cores is given by 
\begin{align}
ds^2&=-A(r, \eta)\,dt^2+\frac{dr^2}{A(r, \eta)}+D^2(r)\,d\Omega^2,\nonumber\\
A(r, \eta)&=1-\frac{2M}{r}\,e^{-\eta/r},\qquad D(r)=r.\label{visser2}    
\end{align}

Against this backdrop, the present paper proposes and analyses in detail the \textit{screened Simpson-Visser} (SSV) geometries, a new static, spherically symmetric regular space-time whose metric function combines both deformation mechanisms without a nonlinear electrodynamics source. The metric function is chosen as,   
\begin{equation}
  A(r, a, \eta)=1-\frac{2M}{\sqrt{r^2+a^2}}\exp\!\left(-\frac{\eta}{\sqrt{r^2+a^2}}\right).\label{function}
\end{equation}
The two dimensionless parameters $a/M$ and $\eta/M$ control, respectively, the wormhole-throat regularization and the exponential screening of the gravitational potential. Setting $\eta=0$ or $a=0$ recovers the two parent models, respectively the metrics (\ref{visser}) or (\ref{visser2}), while $a=\eta=0$ yields the Schwarzschild singular solution.

Alongside the metric construction and the analysis of its global causal structure, we carry out a comprehensive study of the thermodynamic, geodesic, and topological properties of the SSV
geometry.  The Hawking temperature and thermodynamic phase structure are derived and compared with those of the parent models; in particular, a Davies-type phase transition~\cite{Davies1977} in
the specific heat is shown to persist across the two-parameter family.  The shadow radius and photon-sphere radius are computed and confronted with the Event Horizon Telescope images of M87*
and Sgr~A*~\cite{EHTL1, EHTL4, EHTL6, EHTL12, EHTL17}, providing observational bounds on the parameter space. The innermost stable circular orbit (ISCO) and the spectral energy emission rate are analysed as probes of the accretion-disc structure.  Finally, following the topological framework of Cunha and Herdeiro~\cite{CunhaHerdeiro2020} and the extension to spherically symmetric black holes by Wei~\cite{PVPC3} via Duan's $\phi$-mapping topological current theory~\cite{Duan1979,Duan1984}, we compute the winding number of the photon sphere and show that $w = -1$ is a topological invariant of the SSV family.

The remainder of the paper is organized as follows.  Section~\ref{sec:2} introduces the SSV metric and discusses its causal structure and curvature regularity.  Section~\ref{sec:2} analyses the horizon structure via the Lambert $W$ function and the energy conditions. Sections~\ref{sec:3} present the thermodynamic analysis and the phase structure. Section~\ref{sec:4} covers null and timelike geodesics, the photon sphere, shadow radius, and
ISCO.  Section~\ref{sec:5} studies the energy emission rate.  Section~\ref{sec:6} to \ref{sec:7} is devoted to the topological characteristics of the photon sphere and thermodynamics.   Section~\ref{sec:8} collects our conclusions. Throughout this paper, we adopt geometric units with $G=c=1$ and, for numerical calculations, set $M=1$.

\section{Screened SV-Regular Black Hole Solutions}\label{sec:2}

In this section, we proposed a new regular spacetime geometry, referred to as the SSV solution, and analyzed its geometric and physical properties in detail. In particular, we investigated its thermodynamic behavior, geodesic structure, and the corresponding energy emission rate.

We consider a static and spherically symmetric SSV solution. The corresponding line element is expressed as
\begin{equation}
d s^{2}=-A(r, a, \eta) d t^{2}+\frac{d r^{2}}{A(r, a, \eta)}+D^2(r, a)\,d\Omega^2,\label{metric}
\end{equation}
with $A(r, a, \eta)$ is given in (\ref{function}) and \(D(r, a)=\sqrt{r^2+a^2}.\)\\

This black hole metric interpolates between two well-known non-singular models: setting $a=0$ reproduces the solution discussed in \cite{Alex2020}, while $\eta=0$ corresponds to the model introduced in \cite{Simpson2019}. Asymptotically, the metric function $A(r)$ tends to regular charged-like solution, that is, \(A(r) \simeq 1-\frac{2 M}{\sqrt{r^{2}+a^{2}}}+\frac{2 M \eta}{r^2+a^2}\) in contrast to the singular charged-like solution using the metric  (\ref{visser2}) or the metrics discussed in \cite{Culetu2015b,Balart2014}.

\begin{figure}[ht!]
\centering
\includegraphics[width=0.9\linewidth]{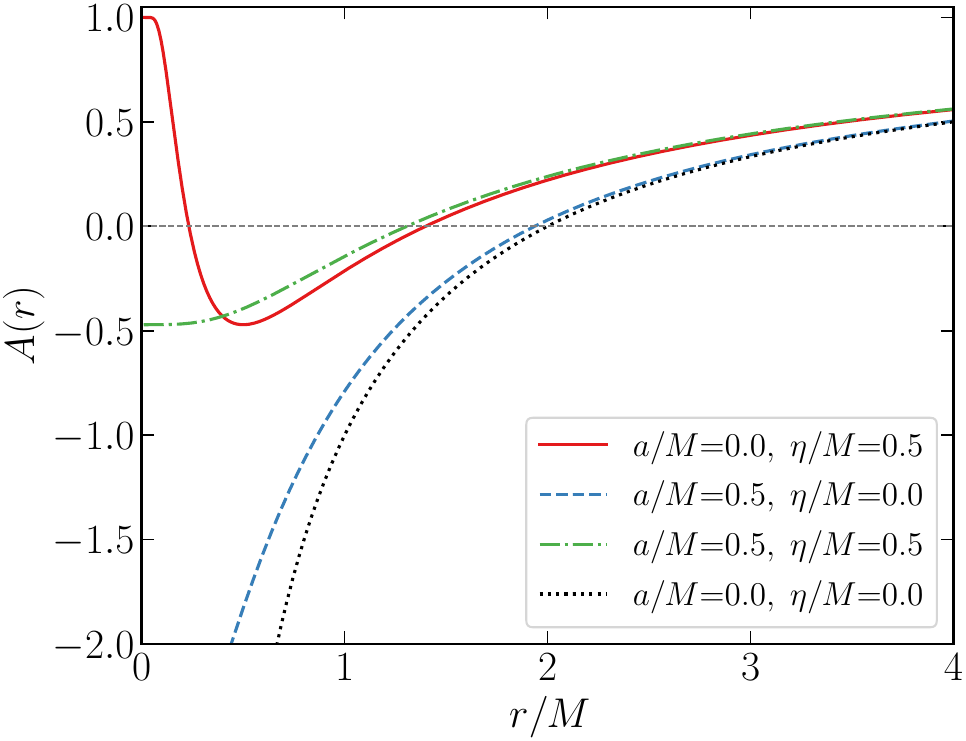}
\caption{Behavior of the metric function $A(r)$ as a function of $r/M$ for four representative parameter combinations: $(a/M, \eta/M) = (0.0, 0.5)$ (red solid), $(0.5, 0.0)$ (blue dashed), $(0.5, 0.5)$ (green dash-dot), and the standard Schwarzschild case $(0.0, 0.0)$ (black dotted).  The regularization parameter $a$ removes the central singularity, replacing it with a regular wormhole-like core, while the screening parameter $\eta$ exponentially suppresses the gravitational potential at all radii.  The combined effect of the two parameters shifts the horizon inward and raises the metric function at the origin.  All curves approach $A(r) \to 1$ at spatial infinity, confirming asymptotic flatness.}
\label{fig:1}
\end{figure}

Figure~\ref{fig:1} illustrates the behavior of the metric function $A(r)$ for four representative combinations of the regularization parameter $a$ and the screening parameter $\eta$.  For the standard Schwarzschild case ($a=\eta=0$, dotted curve), the metric function exhibits the familiar single zero at $r_h = 2M$, marking the event horizon.  When only the screening parameter is switched on ($a=0,\,\eta/M=0.5$, red solid curve), the exponential suppression shifts the horizon inward and lowers the depth of the negative-$A$ region, smoothing the geometry near the origin.  In contrast,
activating only the regularization parameter ($a/M=0.5,\,\eta=0$, blue dashed curve) replaces the curvature singularity at $r=0$ with a finite-area wormhole throat; the metric function reaches a finite negative minimum at $r=0$ rather than diverging.  Finally, the doubly-deformed case ($a/M=0.5,\,\eta/M=0.5$, green dash-dot curve) combines both effects: the horizon shifts, the $r=0$ value is further raised toward zero, and the spacetime is everywhere regular. At large $r$ all curves converge to $A(r)\to 1$, confirming asymptotic flatness.

Analysis of the radial null curves in this metric yields (setting \(ds^2 = 0, d\theta = d\phi = 0\)):
\begin{equation}
    \frac{dr}{dt}=\pm \left(1-\frac{2 M e^{-\frac{\eta}{\sqrt{r^2+a^2}}}}{\sqrt{r^{2}+a^{2}}}\right). 
\end{equation}

It is worth noting that this defines a coordinate speed of light for the metric (\ref{metric}):
\begin{equation}
    v(r)=\left|\frac{dr}{dt}\right|=\left|1-\frac{2 M e^{-\frac{\eta}{\sqrt{r^2+a^2}}}}{\sqrt{r^{2}+a^{2}}}\right|,
\end{equation}
and hence a refractive index of (setting $c=1)$:
\begin{equation}
    n(r)=\left|1-\frac{2 M e^{-\frac{\eta}{\sqrt{r^2+a^2}}}}{\sqrt{r^{2}+a^{2}}}\right|^{-1}.
\end{equation}

\begin{figure}[ht!]
\centering
\includegraphics[width=0.9\linewidth]{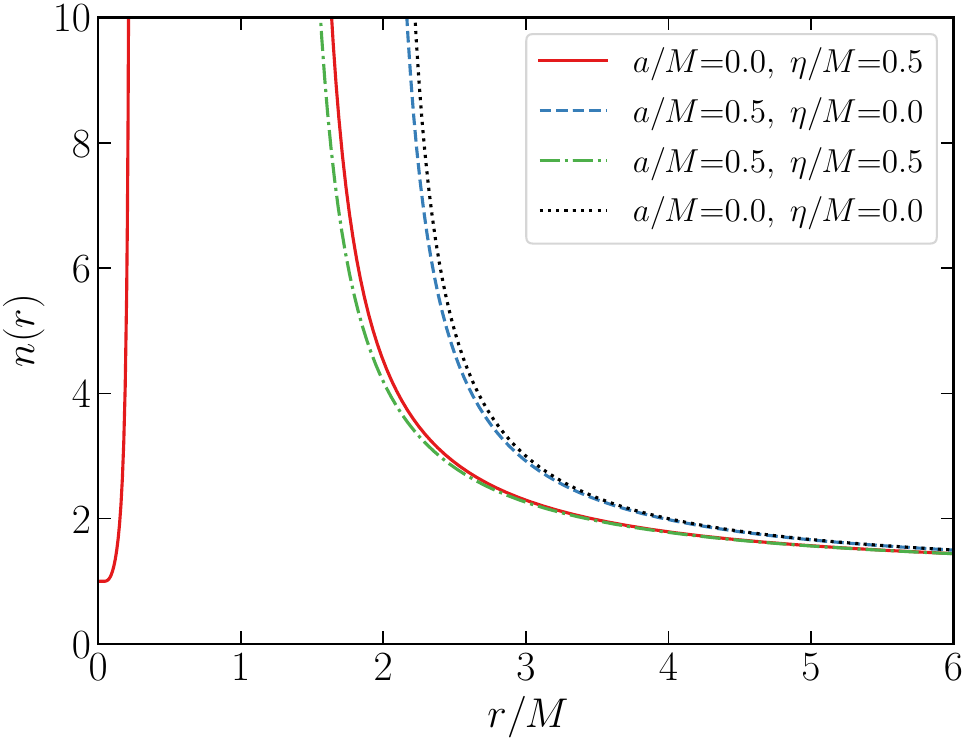}
\caption{Effective refractive index $n(r)$ as a function of $r/M$ for the same four parameter combinations shown in Fig.~\ref{fig:1}.  At large distances $n(r)\to 1$, recovering flat-space optics.  Near the event horizon, $n(r)$ diverges, marking the location where the coordinate speed of light vanishes.  The screening parameter $\eta$ displaces the divergence to smaller radii, while the regularization parameter $a$ softens the near-horizon slope.  The combination of the two parameters yields the smallest effective horizon radius (green dash-dot curve).}
\label{fig:2}
\end{figure}

Figure~\ref{fig:2} displays the refractive index $n(r)$ for the same four parameter sets.  Far from the black hole, $n(r)\to 1$ for all cases, consistent with asymptotic flatness and the recovery of flat-space optics at large distances.  As $r$ decreases toward the horizon, each curve rises steeply, reflecting the increasing gravitational slowing of light.  The divergence of $n(r)$ signals the location of the event horizon, where the coordinate speed of light vanishes.  The screening parameter $\eta$ shifts this
divergence inward, while the regularization parameter $a$ smooths the near-horizon behavior by replacing the sharp $r=0$ singularity with a finite wormhole throat. For the doubly-deformed case ($a/M=0.5,\,\eta/M=0.5$) the horizon is located at the smallest radial coordinate among the four cases, in agreement with the metric function analysis.

At $r=0$, the metric function simplifies as
\begin{equation}
A(0, a, \eta)=\left(1-\frac{2 M e^{-\frac{\eta}{a}}}{a}\right).
\end{equation}

\begin{figure}[ht!]
    \centering
    \includegraphics[width=0.9\linewidth]{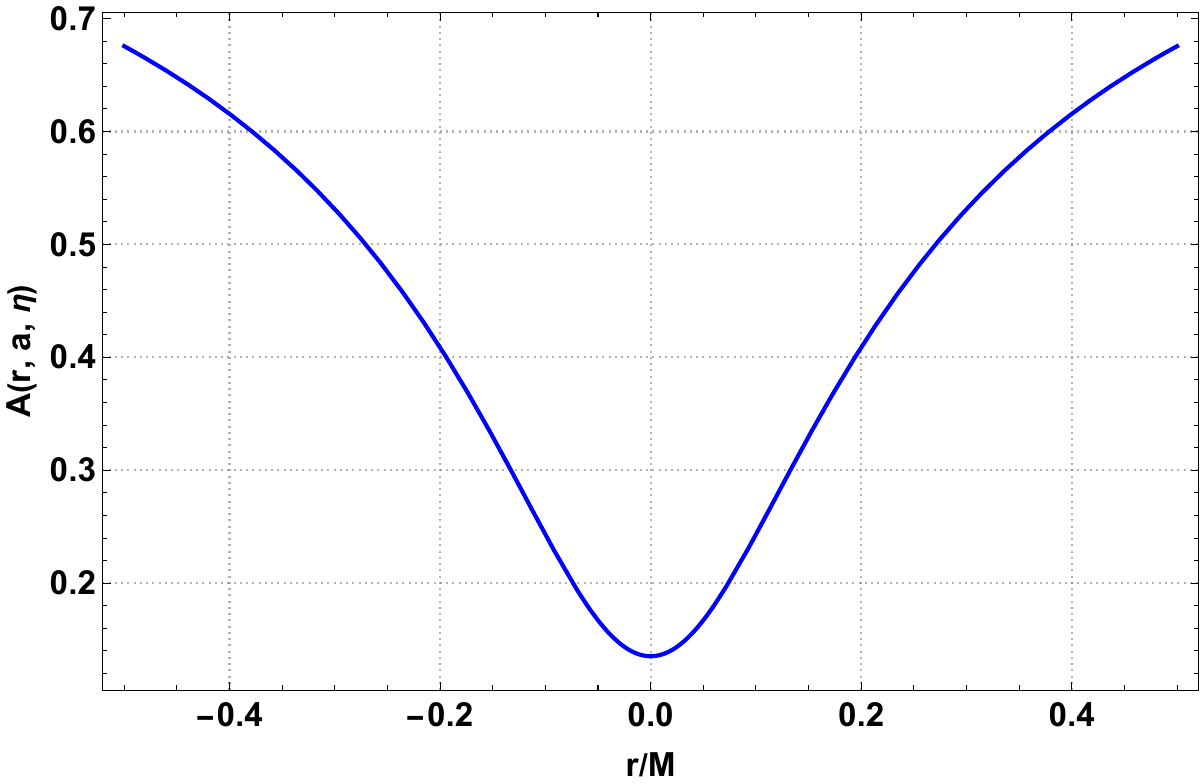}
    \caption{Behavior of the metric function near the origin. Here $a/M=0.1,\,\eta/M=0.2$.}
    \label{fig:function-1}
\end{figure}

Let us now examine the coordinate location(s) of horizon(s) in this geometry:
\begin{itemize}
\item If $a> 2 M e^{-\frac{\eta}{a}}$ for all $r \in \mathbb{R}$, we have $\frac{dr}{dt} \neq 0$, so this geometry is a (two-way) traversable wormhole.

\item If $a= 2 M e^{-\frac{\eta}{a}}$, we have $\frac{dr}{dt}\Big{|}_{r=0} =0$, which indicates this geometry is not a black hole.

\item If $a < 2 M e^{-\frac{\eta}{a}}$, then two horizons exist:

Setting $\rho=\sqrt{r^2+a^2}$ into the metric function yields:
\begin{equation}
A(\rho)=1-\frac{2 M}{\rho}\,e^{-\eta/\rho}.\label{aa3}
\end{equation}

The event horizon $\rho_h$ is determined by
\begin{equation}
A(\rho_h)=0\Rightarrow 1-\frac{2 M}{\rho_h}\,e^{-\eta/\rho_h}=0.\label{aa4}
\end{equation}
The horizon radius following the procedure adopted in \cite{Balart2014} is therefore
\begin{align}
\rho_h (\mbox{outer})&=-\eta\,\left[W\left(0,\,-\frac{\eta}{2M}\right)\right]^{-1},\nonumber\\
\rho_h (\mbox{inner})&=-\eta\,\left[W\left(-1,\,-\frac{\eta}{2M}\right)\right]^{-1},\label{aa5}
\end{align}
where $W$ is the Lambert function, defined by $W(x)\, e^{W(x)} = x$. The physical horizon is given by
\begin{equation}
r_h=\sqrt{\rho_h^2-a^2}.\label{aa6}
\end{equation}
Real horizons ($r_h>0$) exist provided the Lambert function $W$ is real, which implies the constraint
\begin{equation}
-\frac{\eta}{2 M} \geq -\frac{1}{e} \Rightarrow \eta \leq 2 M e^{-1}.\label{aa7}
\end{equation}  
\end{itemize}

If we compare the metric function $A(r, a, \eta)=1-\frac{2 m(r)}{r}$, then the mass profile is given by
\begin{equation}
m(r)=\frac{r \exp\left[-\frac{\eta}{\sqrt{r^2+a^2}}\right]}{\sqrt{r^2+a^2}}\,M.\label{mass}
\end{equation}

\begin{figure}[ht!]
    \centering
    \includegraphics[width=0.9\linewidth]{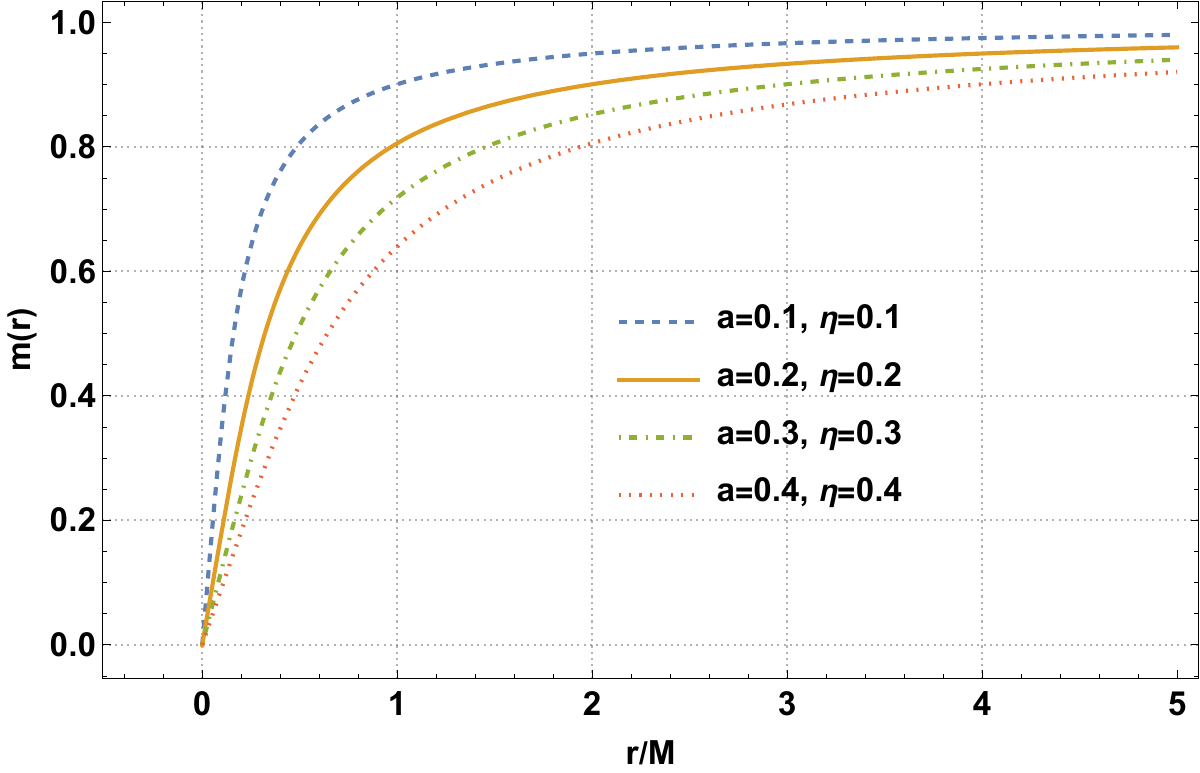}
    \caption{Behavior of the mass profile $m(r)$ as a function of the dimensionless radial distance for various $a$ and $\eta$.}
    \label{fig:mass-profile}
\end{figure}

Near the origin $r \to 0$, we have $m(r) \propto r$, a typical behavior of regular black hole mass functions. A binomial expansion yields
\begin{equation}
m(r)=\frac{M}{\sqrt{1+a^2/r^2}}\left[1-\frac{\eta/r}{\sqrt{1+a^2/r^2}}+\mathcal{O}(\eta/r)^2\right].
\end{equation}
At spatial infinity $r \to \infty$, we find $m(r) \to M$, the ADM mass. Moreover, the density profile $\rho(r)$ for the considered space-time using the mass profile (\ref{mass}) is given by
\begin{align}
    \rho(r)=\frac{m'(r)}{4 \pi (r^2+a^2)}= \frac{M \, e^{-\eta / \sqrt{r^2 + a^2}} \left( a^2 + \frac{\eta \, r^2}{r^2 + a^2} \right)}{4 \pi (r^2 + a^2)^{5/2}}.\label{density}
\end{align}

\begin{figure}[ht!]
    \centering
    \includegraphics[width=0.9\linewidth]{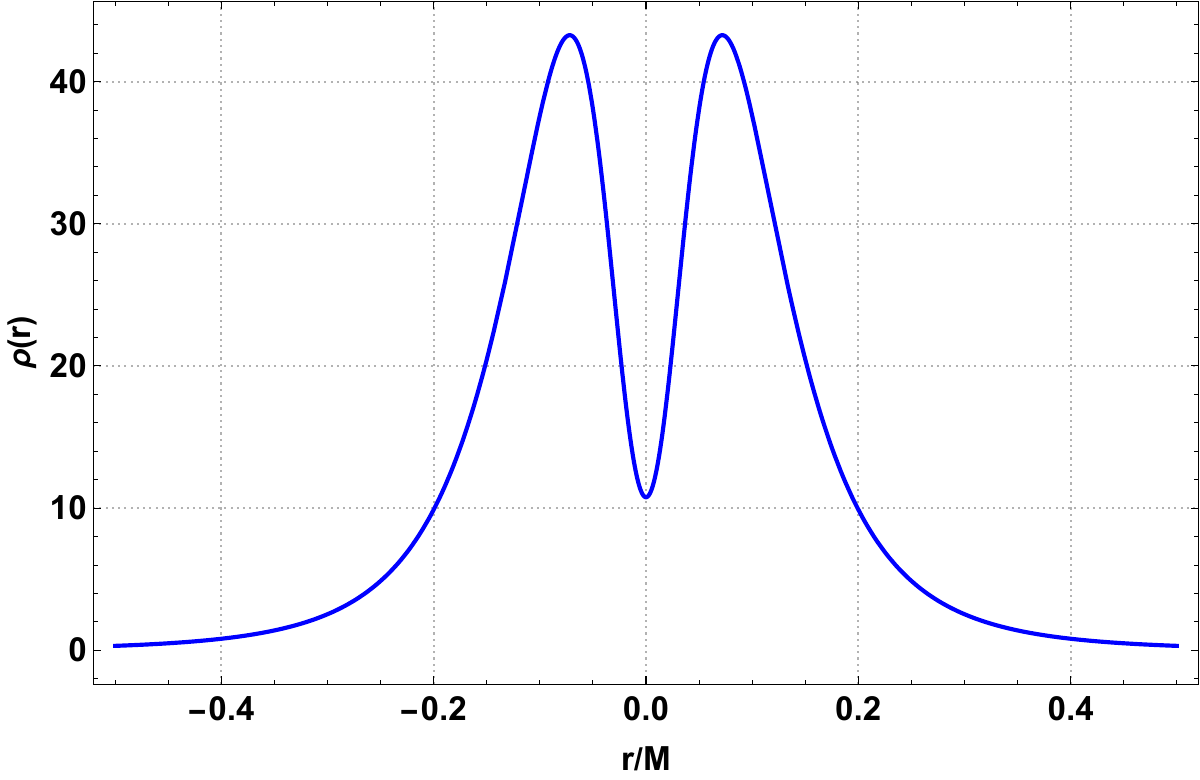}
    \caption{Behavior of the density profile $\rho(r)$ near the origin. Here $a/M=0.1,\,\eta/M=0.2$.}
    \label{fig:density}
\end{figure}

We would require that $\lim_{r \to 0} \rho(r)=\rho_0 \neq 0$ for a de-Sitter core and $\lim_{r \to 0} \rho(r) = 0$ for Minkowski core. In our case at hand, we find that 
\begin{equation}
\lim_{r \to 0} \rho(r)=\frac{M}{4 \pi a^3}\,e^{-\eta/a}=\rho_0. \label{limit}
\end{equation}
This implies that the considered solution is asymptotically de-Sitter Core.

For \(a >0\), the exponential expression factor has the properties:
\begin{equation}
    \lim_{r \to 0^{+}}  e^{-\frac{\eta}{\sqrt{r^2+a^2}}}=e^{-\eta/a}=\lim_{r \to 0^{-}}  e^{-\frac{\eta}{\sqrt{r^2+a^2}}}
\end{equation}
This implies that for any $a>0$, the limit is same on either sides, and hence, no discontinuity occurs at $r=0$.

\subsection{Scalar curvature and Energy Conditions}

For the chosen space-time metric (\ref{metric}), we proceed to compute several important scalar quantities, including the Ricci scalar, the Kretschmann scalar, and the squared Ricci tensor invariants. These scalars provide insight into the geometric and physical properties of the space-time, such as the presence or absence of singularities and the overall curvature structure. Furthermore, we examine the validity of the energy conditions-namely, the weak, strong, and dominant energy conditions-to assess whether the matter content or effective stress-energy tensor associated with this geometry satisfies physically reasonable constraints \cite{chandrasekhar1984,Wald1984}. This analysis is crucial for understanding the physical viability and stability of the space-time configuration under consideration.

We now determine various scalar quantities using the Riemann-Cartan geometry and analyze the outcomes. The squared Ricci scalar ($\mathcal{I}=R_{ab}R^{ab}$), the Kretschmann invariant ($\mathcal{K}=R_{abcd}R^{abcd}$) and the Ricci scalar ($\mathcal{R}=g^{ab}\,R_{ab}$) for the considered space-time (\ref{metric}) are given by 

{\scriptsize
\begin{widetext}
\begin{align}
\mathcal{I}&=\frac{2 e^{-\frac{2 \eta}{\sqrt{a^2 + r^2}}}}{(a^2 + r^2)^7}\Bigg[2 a^{10} e^{\frac{2 \eta}{\sqrt{a^2 + r^2}}} 
+ M^2 r^4 \eta^2 \Big( 8 r^2 + \eta (-4 \sqrt{a^2 + r^2} + \eta) \Big)+ a^8 \Big\{9 M^2 + 6 e^{\frac{2 \eta}{\sqrt{a^2 + r^2}}} r^2\nonumber\\
&- 2 e^{\frac{\eta}{\sqrt{a^2 + r^2}}} M \left(3 \sqrt{a^2 + r^2} + \eta\right) \Big\}+ 2 a^2 M r^2 \eta \Big\{M \sqrt{a^2 + r^2} \eta^2 + M r^2 \left(2 \sqrt{a^2 + r^2} + 3 \eta\right)\nonumber\\
& + e^{\frac{\eta}{\sqrt{a^2 + r^2}}} r^2 (2 r^2 - \sqrt{a^2 + r^2} \eta) \Big\}+ a^6 \Big\{ 6 e^{\frac{2 \eta}{\sqrt{a^2 + r^2}}} r^4 - 12 e^{\frac{\eta}{\sqrt{a^2 + r^2}}} M r^2 \sqrt{a^2 + r^2}+ M^2 \left(18 r^2 + 2 \sqrt{a^2 + r^2} \eta + \eta^2\right) \Big\}\nonumber\\
&+ a^4 r^2 \Big\{2 e^{\frac{2 \eta}{\sqrt{a^2 + r^2}}} r^4 + M^2 \left(9 r^2 + \left(6 \sqrt{a^2 + r^2} - \eta\right) \eta\right)- 2 e^{\frac{\eta}{\sqrt{a^2 + r^2}}} M \left(3 r^2 \left(\sqrt{a^2 + r^2} - \eta\right) + \sqrt{a^2 + r^2} \eta^2\right) \Big\}\Bigg],\\
\mathcal{K}&=\frac{4 \, e^{-\frac{2 \eta}{\sqrt{a^2 + r^2}}}}{(a^2 + r^2)^{15/2}}\Bigg[a^{10} e^{\frac{\eta}{\sqrt{a^2 + r^2}}} \Big(-8 M + 3 e^{\frac{\eta}{\sqrt{a^2 + r^2}}} \sqrt{a^2 + r^2} \Big) + a^8 \Bigl\{-16 e^{\frac{\eta}{\sqrt{a^2 + r^2}}} M r^2+ 9 e^{\frac{2 \eta}{\sqrt{a^2 + r^2}}} r^2 \sqrt{a^2 + r^2} \nonumber\\ 
&
+ M^2 \left(9 \sqrt{a^2 + r^2} - 2 \eta\right) \Bigr\}+ M^2 r^4 \Bigl\{ 12 r^4 (\sqrt{a^2 + r^2} - 2 \eta) 
+ 8 r^2 \left(3 \sqrt{a^2 + r^2} - \eta\right) \eta^2+ \sqrt{a^2 + r^2} \eta^4 \Bigr\}\nonumber\\ 
&+ 2 a^2 M r^4 \Bigl\{2 M r^2 \left(3 \sqrt{a^2 + r^2} - 7 \eta\right) 
+ M \left(7 \sqrt{a^2 + r^2} - 3 \eta\right) \eta^2 
+ e^{\frac{\eta}{\sqrt{a^2 + r^2}}} \left(4 r^4 - 2 r^2 \sqrt{a^2 + r^2} \eta\right) \Bigr\}\nonumber\\
& + a^6 \Bigl\{ 9 e^{\frac{2 \eta}{\sqrt{a^2 + r^2}}} r^4 \sqrt{a^2 + r^2} 
- 4 e^{\frac{\eta}{\sqrt{a^2 + r^2}}} M r^2 \sqrt{a^2 + r^2} \eta 
+ M^2 \left(\sqrt{a^2 + r^2} \eta^2 + 2 r^2\left(3 \sqrt{a^2 + r^2} + 8 \eta\right)\right) \Bigr\} \nonumber\\
& + a^4 r^2 \Big\{ 3 e^{\frac{2 \eta}{\sqrt{a^2 + r^2}}} r^4 \sqrt{a^2 + r^2} 
+ 8 e^{\frac{\eta}{\sqrt{a^2 + r^2}}} M r^2 \left(2 r^2 - \sqrt{a^2 + r^2} \eta\right) 
+ M^2 \Big\{\eta^2 \left(-9 \sqrt{a^2 + r^2} + 2 \eta\right)\notag\\& + r^2 \left(-3 \sqrt{a^2 + r^2} + 14 \eta\right) \Big\}\Big\}\Bigg],
\end{align}
\begin{align}
\mathcal{R}&=\frac{e^{-\frac{\eta}{\sqrt{a^2 + r^2}}}\Big[a^4 \Big(6 M - 2 e^{\frac{\eta}{\sqrt{a^2 + r^2}}} \sqrt{a^2 + r^2} \Big)
+ 2 M r^2 \eta^2+ a^2 \Big(6 M r^2 - 2 e^{\frac{\eta}{\sqrt{a^2 + r^2}}} r^2 \sqrt{a^2 + r^2} + 2 M \sqrt{a^2 + r^2} \eta \Big)\Big]}{(a^2 + r^2)^{7/2}}
\end{align}
\end{widetext}
}
From the above scalar quantities, we observe that at $r=0$ these scalar quantities are finite and vanishes at spatial infinity, $r \to \infty$, confirming the regularity of the space-time and asymptotically flat.

\begin{table*}[ht!]
\centering
\begin{tabular}{|c|c|c|}
\hline
\textbf{Energy Condition} & \textbf{Component Form} & \textbf{Physical Meaning} \\
\hline
NEC & $\rho + p_r \ge 0,\;\; \rho + p_{t} \ge 0$ & Energy density for null observers is non-negative \\
\hline
WEC& $\rho \ge 0,\;\; \rho + p_r \ge 0,\;\; \rho + p_{t} \geq 0$ & Energy density for timelike observers is non-negative \\
\hline
SEC& $\rho + p_r + 2 p_{t} \ge 0,\;\; \rho + p_r \ge 0,\;\; \rho + p_{t} \ge 0$ & Matter produces attractive gravity \\
\hline
DEC & $\rho \ge 0,\;\; \rho \ge |p_r|,\;\; \rho \ge |p_{t}|$ & Energy flow is causal, energy density dominates pressure \\
\hline
\end{tabular}
\caption{Various energy conditions in terms of the energy density $\rho$ and principal pressures $(p_r, p_\theta=p_t, p_\phi=p_t)$\cite{Visser1996,Kontou2020,RodriguesMarcos2022}.}
\label{tab:energy-conditions}
\end{table*}

\begin{figure}[ht!]
    \includegraphics[width=0.95\linewidth]{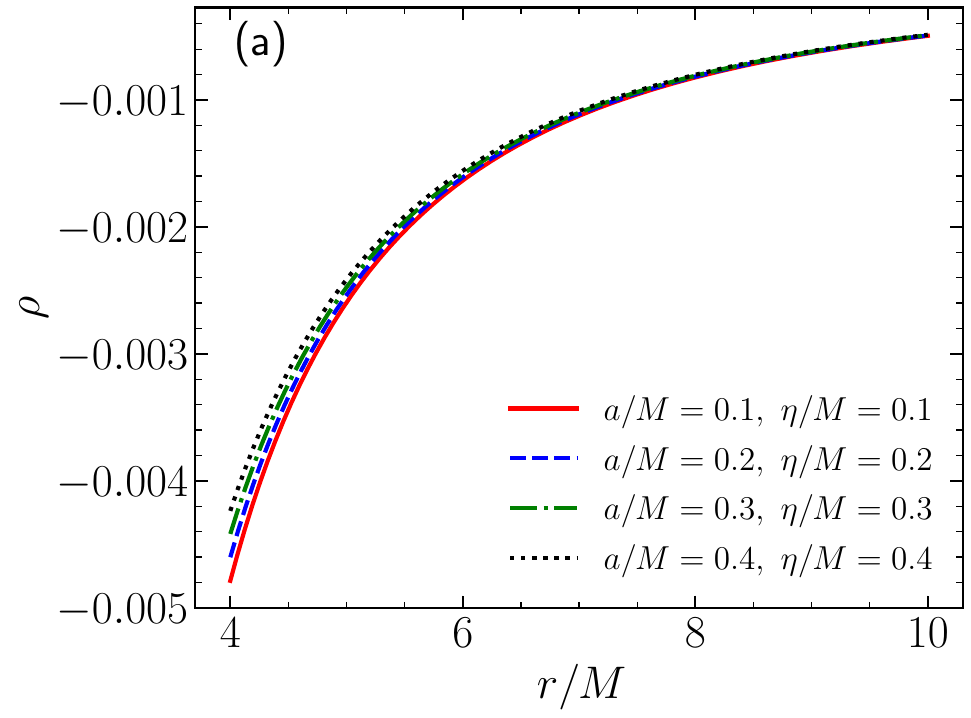}\\
    \includegraphics[width=0.95\linewidth]{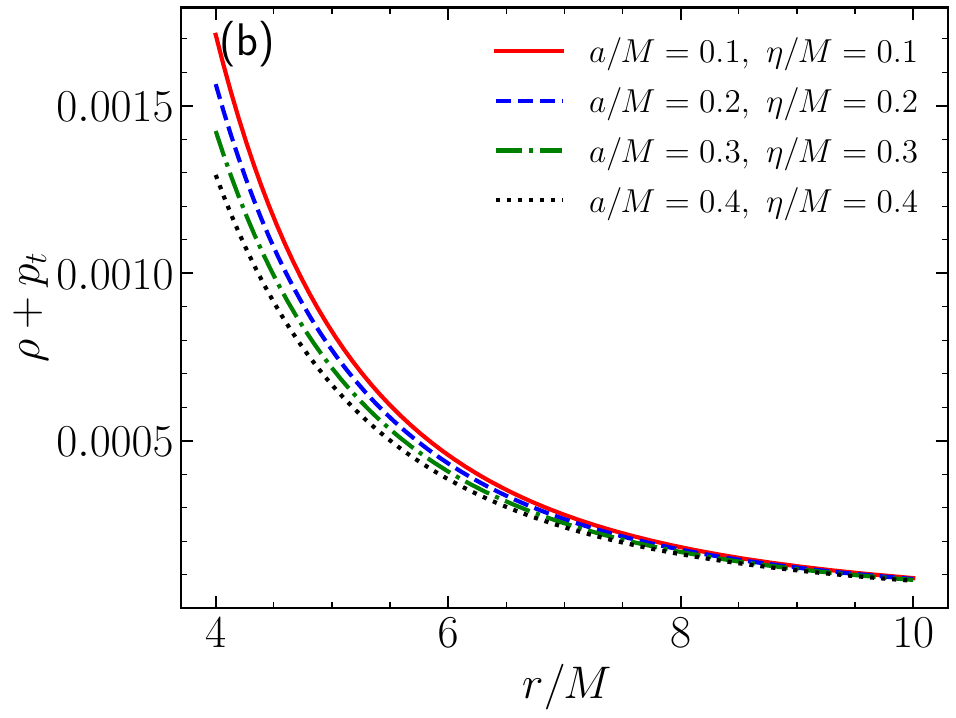}\\
    \includegraphics[width=0.95\linewidth]{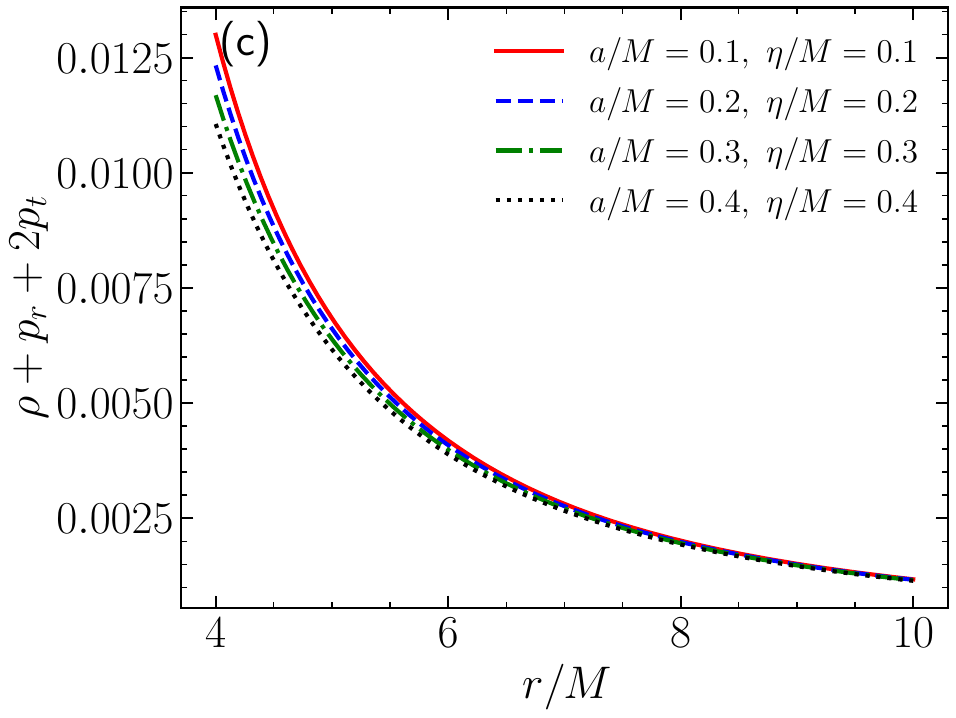}
    \caption{Radial behavior of the energy-condition combinations for the SSV black hole. Panels (a), (b), and (c) show, respectively, the energy density $\rho$, the null-energy-condition combination $\rho+p_t$, and the strong-energy-condition combination $\rho+p_r+2p_t$. These quantities are constructed from the stress--energy tensor components given in Eqs.~(\ref{set1})--(\ref{set3}) and from the combinations summarized in Eq.~(\ref{condition}). The profiles highlight how the effective matter content supporting the regular geometry deviates from the Schwarzschild vacuum, with the largest departures occurring in the near-core region.}
    \label{fig:conditions}
\end{figure}

We now check the energy conditions for the considered regular solutions and analyze the resulting constraints. In the framework of General Relativity, energy conditions are introduced as physically reasonable constraints on matter. They typically impose that the local energy density be non-negative and that the matter flow not exceed the speed of light. The four commonly discussed energy conditions are the weak energy condition (WEC), the strong energy condition (SEC), the dominant energy condition (DEC), and the null energy condition (NEC) \cite{Kontou2020,RodriguesMarcos2022,Visser1996}. It is well known that regular black holes generally violate these energy conditions \cite{Zaslavskii2010}. Since introducing two parameters $a$ and $\eta$ of length dimensions have transformed a singular black hole metric into a regular metric, we must analyze the energy conditions of the obtained black hole solution. 

The Einstein field equations (with $G=1$, $c=1$) are
\begin{equation}
G^{\mu}_{\ \nu}=8\pi T^{\mu}_{\ \nu}.
\end{equation}
For the metric of the form (\ref{metric}) with $A(r, a, \eta)=A$ and $D(r, a)=D$, the mixed Einstein tensor components are:
\begin{align}
G^{t}_{\ t}&
=-\frac{1}{D^{2}}+\frac{2D'^{2}}{D^{2}}+\frac{D''}{D}+\frac{A'D'}{AD},\\
G^{r}_{\ r}&
=-\frac{1}{D^{2}}+\frac{2D'^{2}}{D^{2}}+\frac{D''}{D}+\frac{A'D'}{AD},\\
G^{\theta}_{\ \theta}=G^{\phi}_{\ \phi}&=-\frac{1}{2}\frac{A''}{A}-\frac{1}{4}\frac{A'^{2}}{A^{2}}+\frac{A'D'}{AD}+\frac{D'^{2}}{D^{2}}+\frac{D''}{D}.
\end{align}
Here prime denotes partial derivative w. r to the radial coordinate $r$.

Notice that $G^{t}_{\ t}$ and $G^{r}_{\ r}$ are identical in form; however, in the Einstein equations, they correspond to different components of $T^{\mu}_{\ \nu}$ due to the sign convention.  For a static, spherically symmetric fluid, the stress-energy tensor in mixed form reads
\begin{equation}
T^{\mu}_{\ \nu}=\mathrm{diag}\left(-\rho,\,\,p_{r},\,\,p_{t},\,\,p_{t}\right),
\end{equation}
where $\rho$ is the energy density, $p_{r}$ the radial pressure, and $p_{t}$ the tangential pressure.  Einstein's equations then give:
\begin{align}
8\pi\rho &= -G^{t}_{\ t},\\
8\pi p_{r} &= G^{r}_{\ r},\\
8\pi p_{t} &= G^{\theta}_{\ \theta}=G^{\phi}_{\ \phi}.
\end{align}

The stress-energy tensor components in mixed form are therefore:
\begin{align}
8\pi\rho&= -\frac{1}{r^{2}+a^{2}}\left( \frac{r^{2}}{r^{2}+a^{2}} + \frac{A'}{A}r \right),\label{set1}\\
8\pi p_{r}&= \frac{1}{r^{2}+a^{2}}\left( \frac{r^{2}}{r^{2}+a^{2}} + \frac{A'}{A}r \right) = -8\pi\rho,\label{set2}\\
8\pi p_{t}&= \frac{1}{r^{2}+a^{2}} - \frac{A''}{2A} - \frac{A'^{2}}{4A^{2}} + \frac{A'}{A}\frac{r}{r^{2}+a^{2}},\label{set3}
\end{align}

Now, we have to check various energy conditions listed in \ref{tab:energy-conditions}. Using the above information, we find
\begin{equation}
\left\{
\begin{aligned}
8\pi (\rho+p_r) &= 0,\\
8\pi (\rho+p_{t}) &= \frac{a^2}{(r^2+a^2)^2}-\frac{A''}{2A} - \frac{A'^{2}}{4A^{2}},\\
8\pi (\rho+p_r+2\,p_{t}) &= 16\pi p_{t}.
\end{aligned}
\right.
\label{condition}
\end{equation}
Figure~\ref{fig:conditions} provides a more incisive characterization of the matter sector required to sustain the generalized Simpson--Visser black hole. The three panels show the radial behavior of the energy density $\rho$, the combination $\rho+p_t$, and the quantity $\rho+p_r+2p_t$, which are the combinations relevant for the standard energy-condition analysis. Since Eq.~\eqref{condition} implies $\rho+p_r=0$ identically, any departure from the classical energy conditions is controlled exclusively by the tangential-pressure sector. As a result, the sign of $\rho+p_t$ directly determines whether the null energy condition (NEC) is respected or violated, whereas the sign of $\rho+p_r+2p_t$ governs the corresponding behavior of the strong energy condition (SEC).

As seen in Fig.~\ref{fig:conditions}, these combinations can become negative in the innermost region of the spacetime, signaling that the regular black-hole geometry is supported by an effective exotic matter source. This result is not merely a numerical detail, but rather a structural feature of regular black-hole solutions: avoiding the central singularity generally requires a localized violation of the classical energy conditions. At the same time, all curves decrease in magnitude and tend to zero as $r$ increases, indicating that the matter sector becomes progressively negligible far from the compact object and that the geometry approaches the standard Schwarzschild vacuum limit asymptotically. Figure~\ref{fig:conditions}, therefore, makes transparent that the regularization of the black-hole core is achieved through localized energy-condition violations confined to the near-core region, while the spacetime remains effectively vacuum-like at large distances.

\section{Thermodynamic Properties and Phase Structure}\label{sec:3}

Black hole thermodynamics establishes a deep connection between gravity, quantum theory, and statistical mechanics. A black hole is characterized by thermodynamic quantities such as the Hawking temperature, entropy, and heat capacity, which satisfy laws analogous to those of thermodynamics. The study of these quantities and the associated free energy provides important insights into the phase structure and stability of black holes \cite{Bekenstein1973,Hawking1975,Kubiznak2012}. 

\subsection{Hawking Temperature}

The Hawking Temperature is the temperature associated with the thermal radiation emitted by a black hole due to quantum effects near its event horizon. It was predicted by S. Hawking \cite{Hawking1975}. According to this idea, quantum fluctuations near the event horizon create particle-antiparticle pairs, with one particle falling into the black hole and the other escaping as Hawking Radiation. This radiation makes the black hole behave like a thermal object with a temperature. 

The surface gravity is defined by \cite{Hawking1975,Wald1984,Wald2001}
\begin{equation}
\kappa=\lim_{r \to r_h} \sqrt{-\frac{1}{2}\left(\nabla^{\mu}\xi^{\nu}\right)\,\left(\nabla_{\mu}\xi_{\nu}\right)}=\frac{1}{2} \,A'(r_h),\label{aa8}
\end{equation}
where $\xi^{\mu}$ is the time-like Killing vector field that becomes null on the event horizon.

Using Eq.~(\ref{aa3}), we compute the derivative as follows:
\begin{equation}
\frac{dA}{d\rho}=-2M e^{-\eta/\rho}\,\frac{\eta-\rho}{\rho^3}.\label{aa9}
\end{equation}
As \(\frac{d\rho}{dr}=\frac{r}{\rho}\), we obtain
\begin{equation}
A'(r)=\frac{dA}{d\rho}\,\frac{r}{\rho}=-2M r\, e^{-\eta/\rho}\,\frac{\eta-\rho}{\rho^4}.\label{aa10}
\end{equation}
Evaluating at the horizon and using the horizon condition
\begin{equation}
\frac{2M}{\rho_h}\,e^{-\eta/\rho_h}=1,\label{aa11}
\end{equation}
we find the Hawking temperature given by
\begin{equation}
T=\frac{\kappa}{2\pi}=\frac{r_h}{4\pi \sqrt{r^2_h+a^2}}\left(\frac{1}{\sqrt{r^2_h+a^2}}-\frac{\eta}{r^2_h+a^2}\right).\label{aa12}
\end{equation}

\begin{itemize}
    \item When $a=0$, the Hawking temperature simplifies to
\begin{equation}
T=\frac{1}{4\pi r_h}\left(1-\frac{\eta}{r_h}\right),\quad r_h=\rho_h,\label{aa13}
\end{equation}
where $\rho_h$ is given in (\ref{aa5}).

\item When $\eta=0$, the Hawking temperature simplifies to (now horizon radius $r_h=\sqrt{4 M^2-a^2}$)
\begin{equation}
T=\frac{r_h}{4\pi \left(r^2_h+a^2\right)}=\frac{\sqrt{4 M^2-a^2}}{16\pi M^2}
\end{equation}
which further reduces to the standard Schwarzschild black hole case for $a=0$.
\end{itemize}

From the above analysis, we observe that the Hawking temperature is influenced by the regularization parameters $a$ and $\eta$.

\begin{figure}[ht!]
\centering
\includegraphics[width=0.95\linewidth]{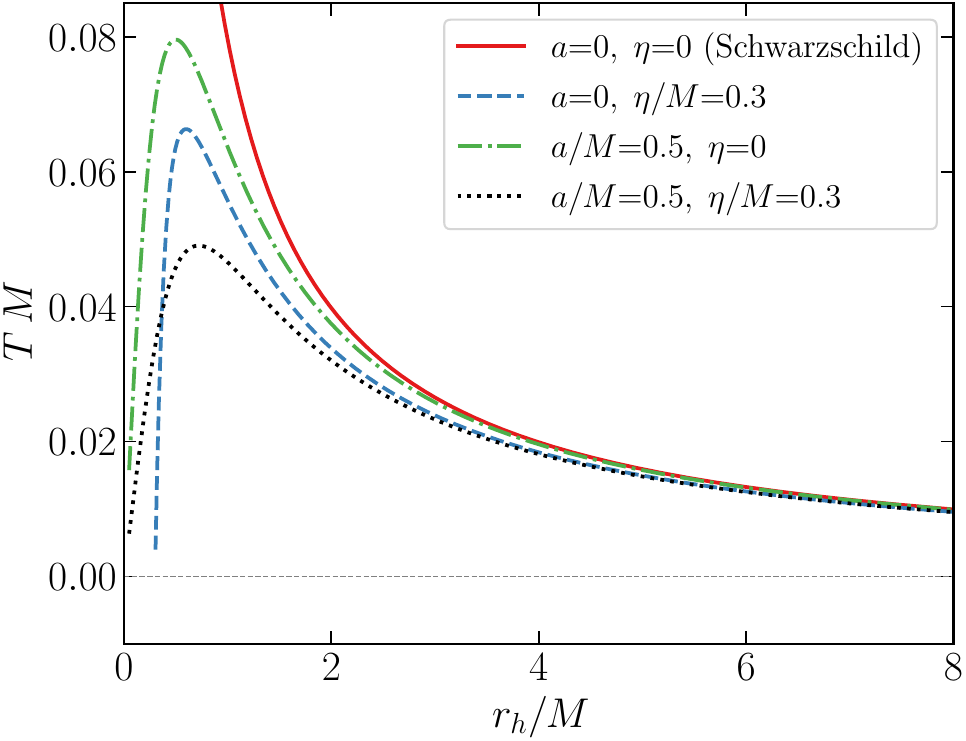}
\caption{Hawking temperature $TM$ as a function of the horizon radius $r_h/M$ for four representative parameter sets: $(a,\,\eta)=(0,0)$ (black dotted, standard Schwarzschild), $(0,\,0.3M)$ (red solid), $(0.5M,\,0)$ (blue dashed), and $(0.5M,\,0.3M)$ (green dash-dot).  Both the regularization parameter $a$ and the screening parameter $\eta$ suppress the Hawking temperature at every horizon radius. The screening parameter introduces an extremal configuration at $r_h = \eta$ where $T=0$, while the regularization parameter shifts the entire curve to larger effective horizon sizes. The combined effect (green dash-dot) produces the most strongly suppressed temperature profile.  All curves tend to zero as $r_h \to \infty$.}
\label{fig:3}
\end{figure}

Figure~\ref{fig:3} shows the Hawking temperature $T$ as a function of the horizon radius $r_h$ for several combinations of $a$ and $\eta$.  For the pure Schwarzschild case ($a=\eta=0$), the
temperature decreases monotonically as $r_h$ increases, following the familiar $T \propto 1/(4\pi r_h)$ law.  Introducing the screening parameter $\eta>0$ (with $a=0$) suppresses $T$ at every
horizon size and introduces a zero-temperature extremal configuration at the inner horizon $r_h = \eta$, where $T=0$; for $r_h < \eta$ the temperature formula yields a negative value, which is unphysical and signals the absence of a well-defined black hole in that regime.  Introducing the regularization parameter $a>0$ (with $\eta=0$) similarly reduces $T$ and shifts the peak of the
temperature curve to larger $r_h$, because the effective radial scale is now $\rho_h = \sqrt{r_h^2+a^2} > r_h$.  When both parameters are present ($a/M=0.5,\,\eta/M=0.3$), these two effects combine: the temperature is further suppressed, the allowed range of $r_h$ is narrowed, and the extremal radius is pushed outward.  All four curves converge to $T\to 0$ as $r_h \to \infty$, consistent with the asymptotic flatness of the geometry.

\subsection{Entropy}

Before proceeding to the system's entropy, let us first find the horizon area. It is given by
\begin{equation}
\mathcal{A}=\int \int \sqrt{g_{\theta\theta}\,g_{\phi\phi}} d\theta d\phi=4\pi (r_h^2+a^2).\label{area}
\end{equation}
The horizon area does not explicitly depend on the screening parameter $\eta$, but $\eta$ influences the horizon radius $r_h$.  Consequently, the area of the horizon also changes.

The entropy consistent with the first law of black hole thermodynamics can be obtained from
\begin{equation}
S = \int \frac{dM}{T}.
\end{equation}
Evaluating the above integral yields
\begin{equation}
S=2\pi \int \sqrt{r^2+a^2}\,e^{\frac{\eta}{\sqrt{r_h^2+a^2}}}dr.
\label{entropy1}
\end{equation}
The integral (\ref{entropy1}) does not have a simple closed form.  A series expansion in $\eta$ gives
\begin{align}
  S& \approx \pi\Bigg[
r_h\sqrt{r_h^2+a^2}
+(a^2+\eta^2 )\ln\!\left(r_h+\sqrt{r_h^2+a^2}\right)\notag\\&+\eta \left( 2\, r_h +\frac{1}{3a}\eta^2 \arctan\!\left(\frac{r_h}{a}\right)\right)
\Bigg].
\label{entropy2}
\end{align}

In the limiting case \(\eta \rightarrow 0\), corresponding to the model introduced in~\cite{Simpson2019}, the standard entropy of the SV black hole and the area relation are recovered.  Furthermore, for $a=0=\eta$ the expression reduces to the standard Schwarzschild black hole case $S_{\rm Sch}=\pi r_h^2$.

Because Eq.~(\ref{entropy2}) contains a logarithmic term
$\ln(r_h+\sqrt{r_h^2+a^2})$ that diverges as $r_h\to 0$, the expression carries an undetermined additive constant, the integration constant of the original indefinite integral~(\ref{entropy1}).  Throughout the analysis below, we normalize each curve by subtracting its value at a small reference radius $r_h^{\min}=0.05\,M$, so that the physically meaningful differences
between curves are preserved while the divergent offset is removed.

\begin{figure}[ht!]
\centering
\includegraphics[width=0.95\linewidth]{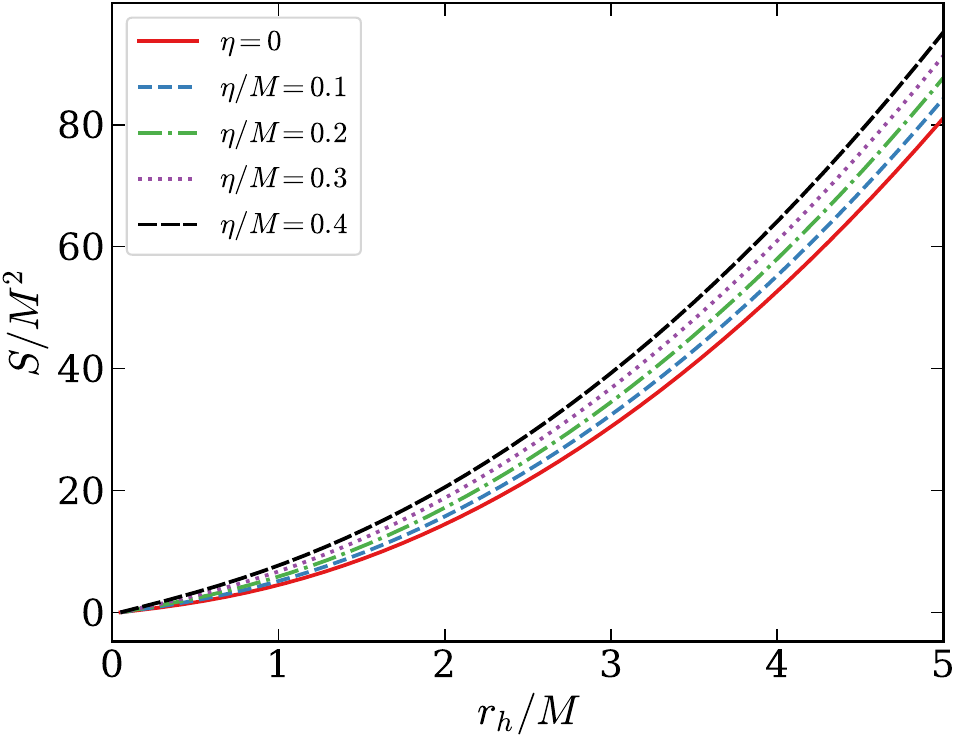}
\caption{Entropy $S/M^{2}$ from Eq.~(\ref{entropy2}) as a function of
the horizon radius $r_{h}/M$ for fixed $a/M=0.5$ and five values of the screening parameter: $\eta=0$ (red solid), $\eta/M=0.1$ (blue dashed), $\eta/M=0.2$ (green dash-dot), $\eta/M=0.3$ (purple dotted), and $\eta/M=0.4$ (black dash-dot-dot).  Each curve is shifted so that $S(r_{h}^{\min})=0$, removing the logarithmic integration constant; the physically relevant quantity is the variation of $S$ with $r_{h}$. For $\eta=0$ the entropy reduces to the standard Simpson--Visser result, growing monotonically with the horizon radius in accordance with the second law of black hole thermodynamics. Increasing $\eta$ introduces additional positive contributions through the linear term $2\pi\eta r_{h}$ and the logarithmic correction $\pi\eta^{2}\ln(r_{h}+\sqrt{r_{h}^{2}+a^{2}})$ in Eq.~(\ref{entropy2}), so curves with larger $\eta$ lie uniformly above
those with smaller $\eta$ at every fixed $r_{h}$.  All curves grow
monotonically with $r_{h}$, confirming consistency with the second law throughout the parameter space.}
\label{fig:4a}
\end{figure}

\begin{figure}[ht!]
\centering
\includegraphics[width=0.9\linewidth]{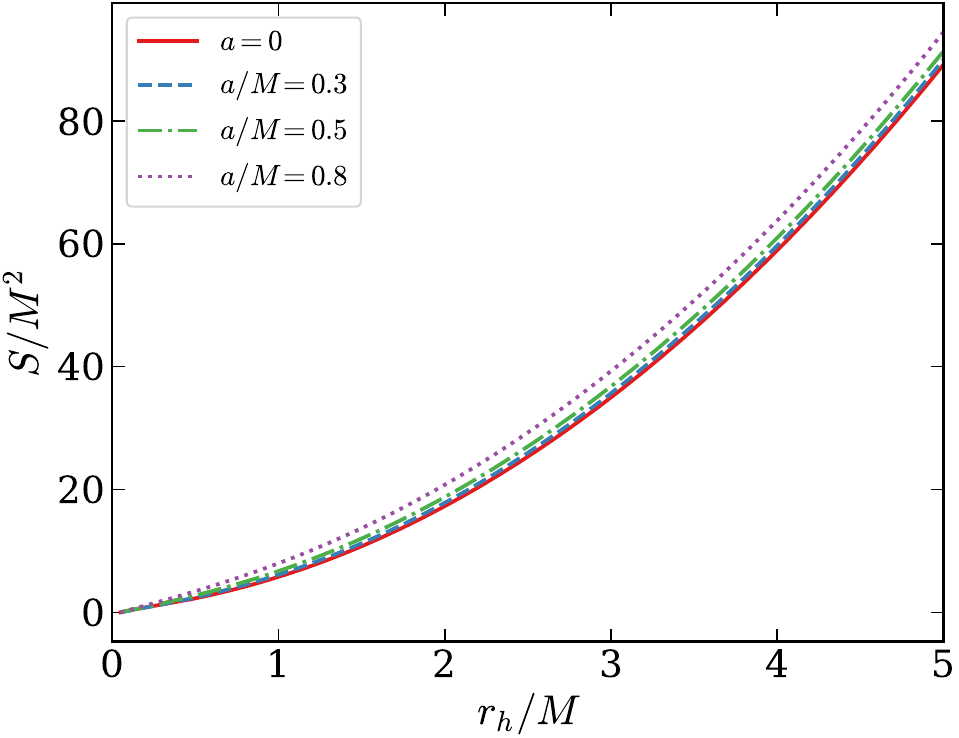}
\caption{Entropy $S/M^{2}$ from Eq.~(\ref{entropy2}) as a function of
$r_{h}/M$ for fixed $\eta/M=0.3$ and four values of the regularization parameter: $a=0$ (red solid), $a/M=0.3$ (blue dashed), $a/M=0.5$ (green dash-dot), and $a/M=0.8$ (black dotted).  As in Fig.~\ref{fig:4a}, each curve is normalized by subtracting its value at $r_{h}^{\min}=0.05\,M$. Increasing $a$ lifts every entropy curve through the geometric contribution encoded in the leading term
$\pi r_{h}\sqrt{r_{h}^{2}+a^{2}}$ of Eq.~(\ref{entropy2}), which
reflects the enlargement of the minimum horizon area by the wormhole
throat of radius $a$.  The coefficient of the logarithmic term
$(a^{2}+\eta^{2})\ln(r_{h}+\sqrt{r_{h}^{2}+a^{2}})$ also grows with
$a$, mildly enhancing the slope of each curve.  At large $r_{h}$ all
curves converge toward the same quadratic growth $S\sim\pi r_{h}^{2}$, since the throat contribution $\pi a^{2}$ becomes sub-leading. The monotonic increase with both $r_{h}$ and $a$ is consistent with the second law of thermodynamics and with the interpretation of $a$ as a geometric regulator that replaces the central singularity with a finite-area wormhole throat.}
\label{fig:4b}
\end{figure}

\begin{figure}[ht!]
\centering
\includegraphics[width=0.95\linewidth]{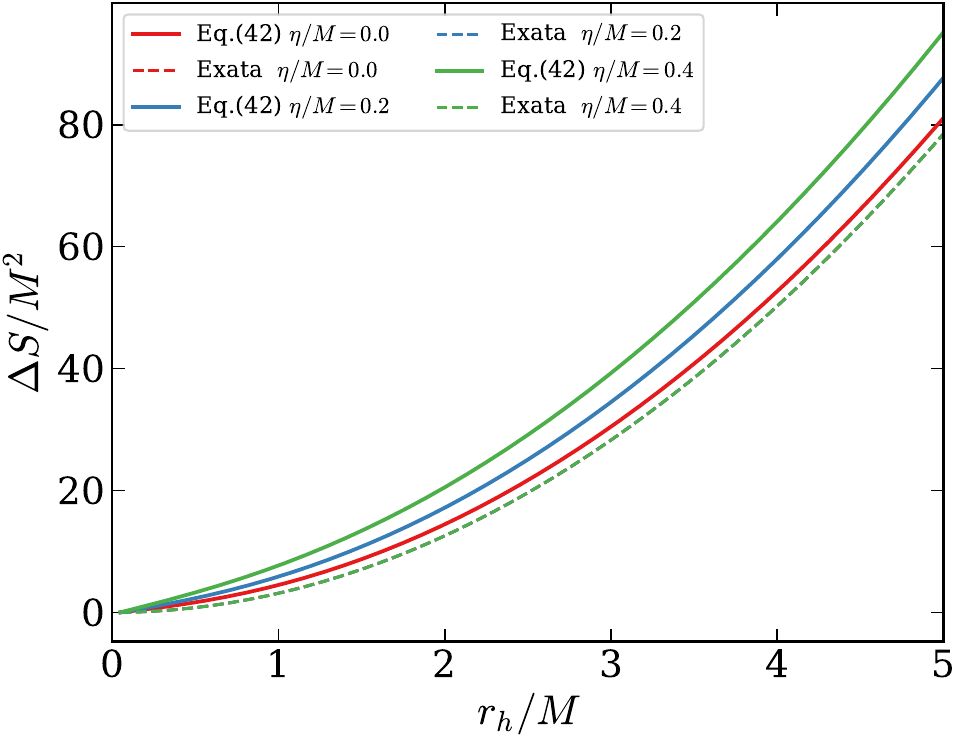}
\caption{Comparison between the entropy from the series expansion
Eq.~(\ref{entropy2}) (solid lines) and the exact Bekenstein--Hawking area law $S_{\rm exact}=\pi(r_{h}^{2}+a^{2})$ [Eq.~(\ref{area}), dashed lines] for $a/M=0.5$ and three values of the screening parameter: $\eta=0$ (red), $\eta/M=0.2$ (blue), and $\eta/M=0.4$ (green).  Both sets of curves are normalized by subtracting their respective values at $r_{h}^{\min}=0.05\,M$.  For $\eta=0$ the two expressions coincide identically, solid and dashed red curves overlap, confirming that Eq.~(\ref{entropy2}) reduces exactly to the area law when screening is absent. As $\eta$ increases, the series-expanded entropy (solid) grows faster than the area law (dashed): the gap widens with $r_{h}$ and is larger for larger $\eta$, measuring the $\eta$-dependent corrections captured by Eq.~(\ref{entropy2}) but absent from the leading-order area formula.  For $\eta/M\lesssim 0.2$ and $r_{h}/M\lesssim 3$ the two descriptions differ by only a few percent, validating the area law as a reliable approximation in that regime; for larger $\eta$ or larger $r_{h}$ the full integral~(\ref{entropy1}) must be evaluated numerically to obtain accurate entropy values.}
\label{fig:4c}
\end{figure}

We present the entropy of Eq.~(\ref{entropy2}) in three complementary figures. Figure~\ref{fig:4a} fixes the regularization parameter and varies the screening parameter $\eta$; Figure~\ref{fig:4b} fixes $\eta$ and varies $a$; and Figure~\ref{fig:4c} directly compares the series approximation
Eq.~(\ref{entropy2}) with the exact Bekenstein--Hawking area law
$S_{\rm exact}=\pi(r_h^2+a^2)$ [cf.\ Eq.~(\ref{area})] for representative
values of $\eta$.

Figure~\ref{fig:4a} displays the entropy from Eq.~(\ref{entropy2}) keeping $a/M=0.5$ fixed and varying $\eta/M$ from $0$ to $0.4$.  For $\eta=0$ the expression reduces to the Simpson--Visser entropy, which grows as $S\approx\pi r_h^2$ for $r_h\gg a$.  Switching on $\eta$ adds two correction terms: a linear contribution $2\pi\eta r_h$ that raises the entropy uniformly at every horizon size, and a logarithmic correction $\pi\eta^2\ln(r_h+\sqrt{r_h^2+a^2})$ that grows slowly with $r_h$. The net effect is a systematic upward shift of the entropy curve that increases monotonically with $\eta$.  Physically, the exponential screening enlarges the accessible thermodynamic phase space, so a screened black hole carries more
entropy than its unscreened counterpart at the same horizon radius.

Figure~\ref{fig:4b} shows the same entropy formula with $\eta/M=0.3$ fixed and $a/M$ varying from $0$ to $0.8$.  Increasing $a$ lifts every curve through the geometric contribution of the wormhole throat: the leading term $\pi r_h\sqrt{r_h^2+a^2}$ in Eq.~(\ref{entropy2}) introduces a positive offset $\pi a^2$ to the entropy relative to the $a=0$ baseline. The logarithmic coefficient $(a^2+\eta^2)$ also grows with $a$, producing a mild enhancement of the entropy slope at intermediate $r_h$.  At large $r_h$ all curves converge to the same quadratic growth $S\sim\pi r_h^2$, since the
throat contribution becomes sub-leading.  The overall picture is consistent with the second law: at fixed $\eta$, the entropy is a monotonically increasing function of both $r_h$ and $a$.

Figure~\ref{fig:4c} places Eq.~(\ref{entropy2}) and the exact area law $S_{\rm exact}=\pi(r_h^2+a^2)$ side by side for $a/M=0.5$ and three values of $\eta$.  When $\eta=0$, the two expressions are identical by construction: all correction terms in Eq.~(\ref{entropy2}) vanish and the leading term $\pi r_h\sqrt{r_h^2+a^2}$ reduces to the area law.  For $\eta>0$ the series formula lies above the area law at every $r_h$, with the gap widening as $r_h$ grows and as $\eta$ increases.  This systematic excess arises from the $\eta$-dependent corrections in Eq.~(\ref{entropy2}), the linear term $2\pi\eta r_h$ and the logarithmic term $\pi\eta^2\ln(\cdots)$, which represent genuine thermodynamic contributions of the exponential screening
that go beyond the geometric area of the horizon.  The comparison therefore quantifies the regime of validity of the area-law approximation: for $\eta/M\lesssim 0.2$ and $r_h/M\lesssim 3$ the two descriptions agree to within a few percent, while for larger values the full integral~(\ref{entropy1}) must be evaluated numerically.

\subsection{Free Energy Analysis}

For screened SV regular black holes, the free energy can be constructed using the thermodynamic quantities derived above.  The free energy is defined as
\begin{align}
F = M - T S, 
\label{free-energy}
\end{align}
where all quantities have their usual meanings.

\begin{figure}[ht!]
\centering
\includegraphics[width=0.9\linewidth]{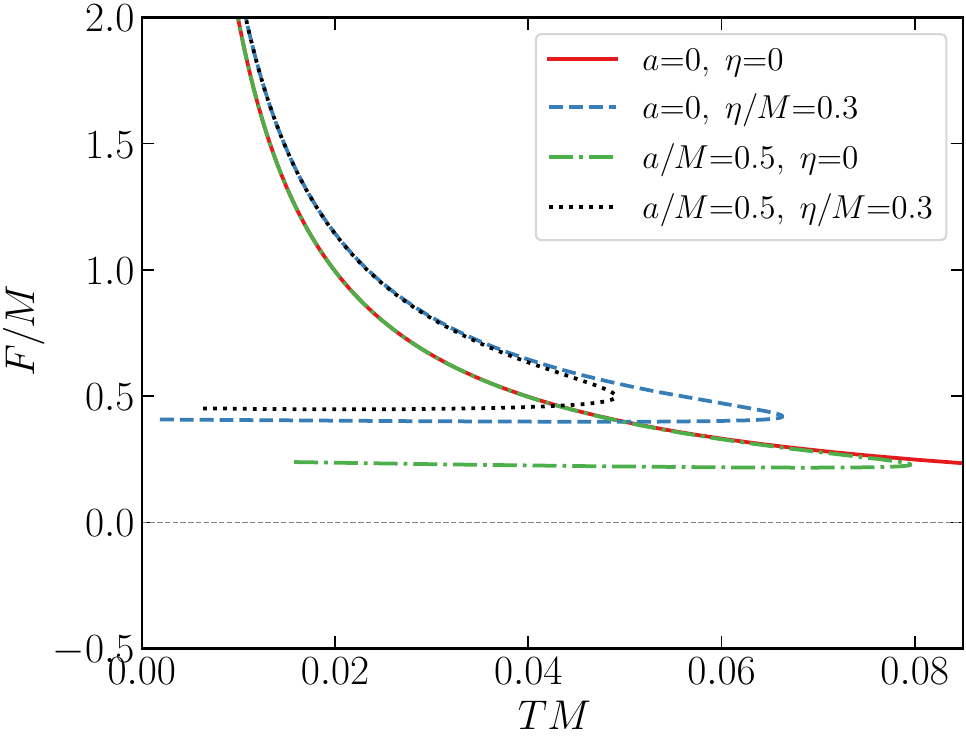}
\caption{Helmholtz free energy $F/M$ as a function of the Hawking temperature $TM$ for the parameter sets $(a,\,\eta)=(0,0)$ (black dotted), $(0,\,0.3M)$ (red solid), $(0.5M,\,0)$ (blue dashed), and $(0.5M,\,0.3M)$ (green dash-dot).  A positive free energy indicates a thermodynamically stable black hole phase; $F$ decreasing toward
zero or negative values signals an approach to a thermal radiation-dominated phase. Both $a$ and $\eta$ restrict the allowed temperature range and increase the free energy at any given temperature, suggesting enhanced thermodynamic stability of the SSV geometry relative to the Schwarzschild benchmark.}
\label{fig:6}
\end{figure}

Figure~\ref{fig:6} shows the Helmholtz free energy $F/M$ as a function of the Hawking temperature $TM$ for the same four parameter sets.  In a Gibbs-like phase diagram interpretation, a positive free energy ($F>0$) indicates a thermodynamically stable black hole phase, while $F<0$ signals the onset of a preferred thermal radiation phase (Hawking--Page-like transition).  For the standard Schwarzschild case ($a=\eta=0$), the free energy is always positive and decreases with temperature, consistent with the absence of a classical phase transition.  Introducing either $\eta>0$ or $a>0$ shifts the $F$--$T$ curve, modifying both the temperature range accessible to the black hole and the slope of the free energy.  In the doubly-deformed case ($a/M=0.5,\,\eta/M=0.3$), the range of accessible temperatures is most narrowly constrained (consistent with Fig.~\ref{fig:3}), and the free energy remains positive throughout, suggesting greater thermodynamic stability compared to the Schwarzschild geometry.

\subsection{Specific Heat} 

In classical thermodynamics, the specific heat $C$ is given by the standard relation:
\begin{equation}
    C=\frac{dM}{dT}.
\end{equation}
In our case, the heat capacity is given by 
\begin{equation}
    C=\frac{4\pi\left( r_h^2+a^2  \right)^{5/2}}{a^2\left(  \sqrt{r_h^2+a^2}-\eta \right)-r_h^2\left(  \sqrt{r_h^2+a^2}-2\eta \right)}. \label{heatc1}
\end{equation}
\begin{itemize}
    \item When $a=0$, the heat capacity simplifies to
\begin{equation}
C=\frac{-4\pi \,r_h^3}{r_h-2\eta }.\label{heatC33}
\end{equation}

\item When $\eta=0$, the heat capacity simplifies to
\begin{equation}
C=\frac{4\pi \left(r^2_h+a^2\right)^2}{a^2-r_h^2},\label{aa14a}
\end{equation}
which further reduces to the standard Schwarzschild black hole case for $a=0$.
\end{itemize}

\begin{figure}[ht!]
\centering
\includegraphics[width=0.9\linewidth]{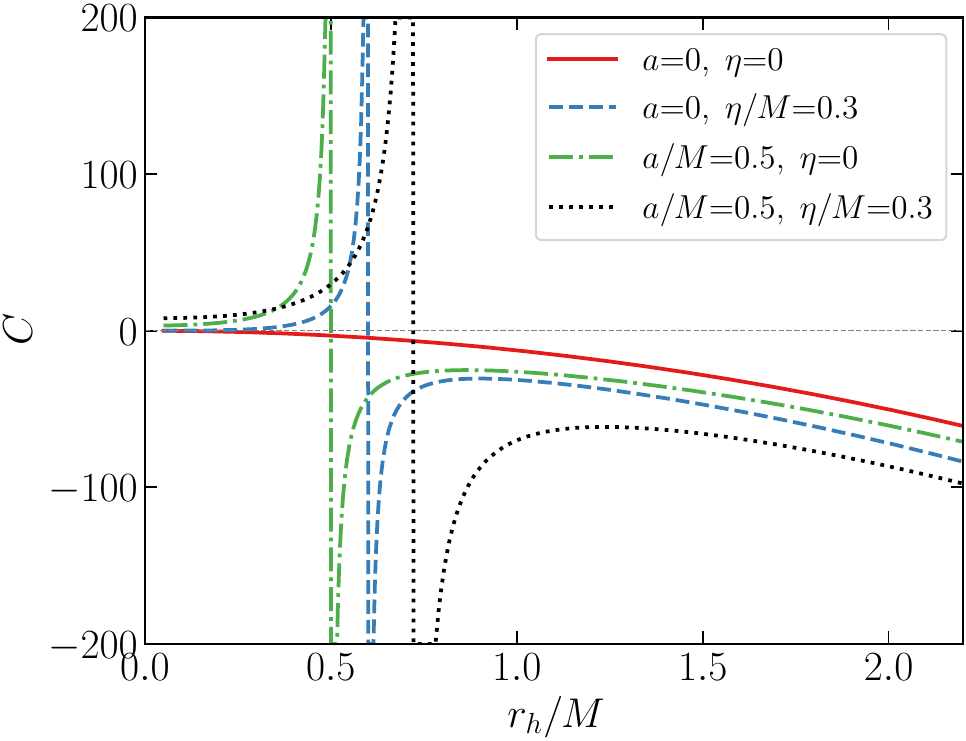}
\caption{Specific heat $C$ as a function of the horizon radius $r_h/M$ for $(a,\,\eta)=(0,0)$ (black dotted), $(0,\,0.3M)$ (red solid), $(0.5M,\,0)$ (blue dashed), and $(0.5M,\,0.3M)$ (green dash-dot).  Positive (negative) values indicate thermodynamically stable (unstable) black hole configurations.  The standard Schwarzschild result ($C<0$, dotted) is recovered in the $a=\eta=0$ limit.  Both $\eta$ and $a$ introduce a Davies-type phase transition at a critical horizon radius where $C$ diverges; for $r_h$ larger than this critical value, the black hole becomes thermally stable ($C>0$).  The doubly-deformed case (green dash-dot) places the transition at the largest critical radius.}
\label{fig:5}
\end{figure}

Figure~\ref{fig:5} displays the specific heat $C$ as a function of $r_h/M$ for the same four representative parameter combinations.  The sign and magnitude of $C$ encode the local thermodynamic stability of the black hole: $C>0$ implies a thermally stable configuration (the black hole heats up when it gains mass), whereas $C<0$ signals an unstable one (the black hole cools as it evaporates).  For the Schwarzschild case ($a=\eta=0$, dotted), $C$ is uniformly negative for all $r_h$, in agreement with the well-known instability of the Schwarzschild black hole.  Introducing
the screening parameter $\eta>0$ (with $a=0$) introduces a divergence in $C$ at a critical horizon radius $r_h^{\rm crit}$, which is the locus where the denominator of Eq.~(\ref{heatc1}) vanishes. This divergence is the hallmark of a Davies-type phase transition \cite{chandrasekhar1984}, separating a region of negative specific heat (small black holes, thermally unstable) from a region of positive specific heat (large black holes, thermally stable).  Increasing $a$ shifts this transition point
to larger $r_h$ and further modifies the asymptotic behavior.  The doubly-deformed case ($a/M=0.5,\,\eta/M=0.3$) exhibits the richest phase structure, with the transition point displaced
to the largest critical radius among the four cases shown.

\section{Photon Sphere, Shadow and ISCO Analysis}\label{sec:4}

The geodesic structure of a spacetime governs the motion of test particles and photons in a gravitational field. Time-like geodesics correspond to the trajectories of massive particles, whereas null geodesics describe the paths of massless particles, such as photons. Analyzing these geodesics provides valuable insights into the causal structure, stability, and observable properties of black holes, including the photon sphere, shadow, light bending, perihelion precession, and accretion disk dynamics. In particular, time-like geodesics allow for the determination of the location of marginally stable circular orbits \cite{chandrasekhar1984,Wald1984}. 

Let us now find the location of both the photon sphere for massless particles and the ISCO for massive particles as functions of $M, a$, and $\eta$.  Due to the spherical symmetry of the metric, one can set $\theta = \pi/2$ without loss of generality and consider the reduced equatorial problem. The Lagrangian function is given by (for related studies see, for example, \cite{Al-Badawi2026,ALBADAWI2025185,Al-Badawi2023,FA5,Faizuddin2026,Ahmed2026} ):
\begin{align}
    \mathcal{L}&=\frac{1}{2} g_{\mu\nu} \frac{dx^{\mu}}{d\tau} \frac{dx^{\nu}}{d\tau}\nonumber\\
    &=\frac{1}{2}\left[-A\left(\frac{dt}{d\tau}\right)^2+\frac{1}{A}\left(\frac{dr}{d\tau}\right)^2+D^2\left(\frac{d\phi}{d\tau}\right)^2\right].\label{ss1}
\end{align}

The Killing vector symmetries lead to the following conserved energy $\mathrm{E}$ and angular momentum $\mathrm{L}$:
\begin{align}
    A \left(\frac{dt}{d\tau}\right)&=\mathrm{E},\label{ss2}\\
    D^2\left(\frac{d\phi}{d\tau}\right)&=\mathrm{L}.\label{ss3}
\end{align}
The radial equation of motion is
\begin{equation}
\left(\frac{dr}{d\tau}\right)^2+V_{\rm eff}=\mathrm{E}^2,\label{ss4}
\end{equation}
where the effective potential for geodesic orbits is
\begin{equation}
    V_{\rm eff}=\left(-\epsilon+\frac{\mathrm{L}^2}{r^2+a^2}\right)\,A(r).\label{ss5}
\end{equation}
Here $\epsilon=0$ for photons and $\epsilon=-1$ for massive particles.

\begin{table*}[tbhp]
\centering
\begin{tabular}{|c|c|c|c|c|c|}
\hline
$a/M (\downarrow) \backslash \eta/M (\rightarrow)$ & 0.1 & 0.2 & 0.3 & 0.4 & 0.5 \\
\hline
0.1 & {2.86156, 5.01886} & {2.71706, 4.83231} & {2.56283, 4.63433} & {2.3958, 4.42176} & {2.21082, 4.18954} \\
0.2 & {2.85632, 5.01886} & {2.71154, 4.83231} & {2.55697, 4.63433} & {2.38953, 4.42176} & {2.20403, 4.18954} \\
0.3 & {2.84755, 5.01886} & {2.7023, 4.83231} & {2.54718, 4.63433} & {2.37905, 4.42176} & {2.19266, 4.18954} \\
0.4 & {2.83523, 5.01886} & {2.68932, 4.83231} & {2.5334, 4.63433} & {2.36429, 4.42176} & {2.17663, 4.18954} \\
0.5 & {2.81931, 5.01886} & {2.67253, 4.83231} & {2.51557, 4.63433} & {2.34518, 4.42176} & {2.15586, 4.18954} \\
\hline
\end{tabular}
\caption{Numerical values of photon sphere and shadow radii $(r_{\rm ph}/M\,,\,R_{\rm sh}/M)$ for various $a$ and $\eta$.}
\label{tab:Photon-Shadow-radius}
\end{table*}

\subsection{Photon Dynamics}

For null geodesics, the effective potential simplifies to
\begin{equation}
    V^{\rm null}_{\rm eff}=\frac{\mathrm{L}^2}{r^2+a^2}\,A(r).\label{null-potential}
\end{equation}

\begin{figure}[ht!]
\centering
\includegraphics[width=0.95\linewidth]{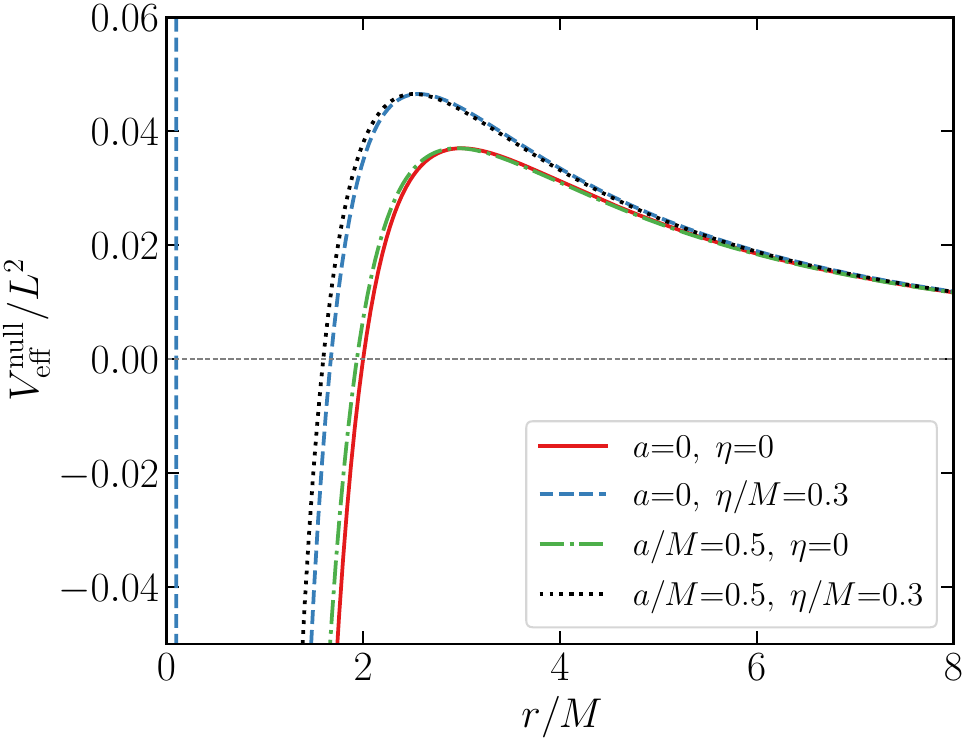}
\caption{Null effective potential $V^{\rm null}_{\rm eff}/L^2$ as a function of $r/M$ for $(a,\,\eta)=(0,0)$ (black dotted), $(0,\,0.3M)$ (red solid), $(0.5M,\,0)$ (blue dashed), and $(0.5M,\,0.3M)$ (green dash-dot).  The maximum of each curve marks the unstable circular photon orbit (photon sphere).  The screening parameter $\eta$ shifts the photon sphere inward and lowers the barrier height, while the regularization parameter $a$ shifts it slightly outward.  The combined effect produces the most inward-displaced and lowest-barrier photon sphere.  The fall-off of the potential at small $r$ reflects the regular geometry of the spacetime near the origin.}
\label{fig:7}
\end{figure}

Figure~\ref{fig:7} shows $V^{\rm null}_{\rm eff}/L^2$ as a function of $r/M$ for the four canonical parameter sets.  Each curve exhibits a pronounced maximum at the photon sphere radius $r_{\rm ph}$, which is the unstable circular orbit of photons.  For the Schwarzschild case ($a=\eta=0$) this maximum occurs at $r_{\rm ph}=3M$ (recoverable from the $\eta=0$ limit of Eq.~(\ref{ss6}) below).  Activating $\eta$ (with $a=0$) lowers and shifts the peak inward, reflecting the exponential screening of the gravitational potential.  Introducing $a$ (with $\eta=0$) moves the maximum outward slightly and raises the barrier height because the angular momentum term $L^2/(r^2+a^2)$ is reduced at any fixed $r$.  The combined effect ($a/M=0.5,\,\eta/M=0.3$) produces the lowest and most inward-displaced peak.  The rapid decrease of $V^{\rm null}_{\rm eff}$ inside the peak signals the capture of photons
by the black hole, while the slow decay at large $r$ reflects the gravitational binding of the photon orbit.

\begin{figure*}[ht]
\centering
\includegraphics[width=0.9\linewidth]{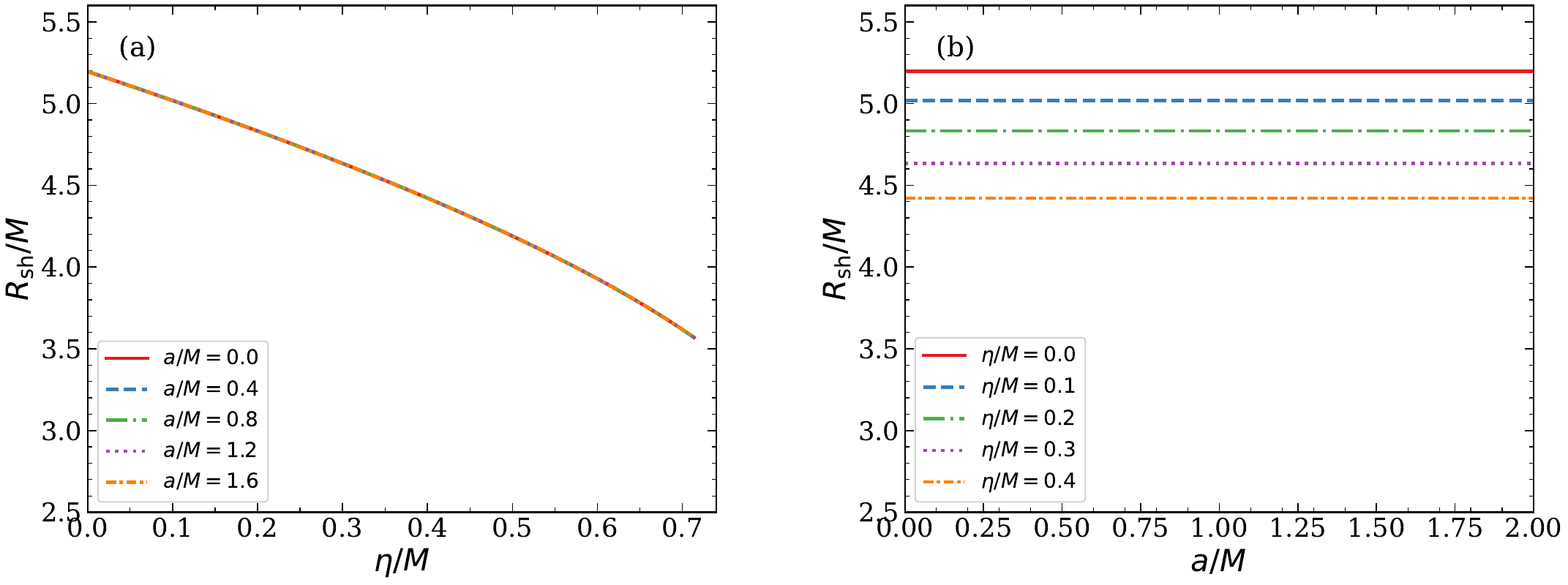}
\caption{Shadow radius $R_{\rm sh}/M$ as a function of the model parameters for the SSV black hole.
Panel~(a): $R_{\rm sh}/M$ vs.\ the screening parameter $\eta/M$ for fixed values $a/M=0.0$ (red solid), $0.4$ (blue dashed), $0.8$ (green dash-dot), $1.2$ (purple dotted), and $1.6$ (orange dash-dot-dot). Panel~(b): $R_{\rm sh}/M$ vs. the regularization parameter $a/M$ for fixed values $\eta/M=0.0$ (red solid), $0.1$ (blue dashed), $0.2$ (green dash-dot), $0.3$ (purple dotted), and $0.4$ (orange dash-dot-dot). In both panels, $R_{\rm sh}$ decreases monotonically with $\eta$, reflecting the reduced photon capture cross-section due to the exponential screening of the gravitational potential, while increasing monotonically with $a$, since a larger wormhole throat enlarges the effective photon orbit.  The upper bound $\eta < 2M/e$ [Eq.~(\ref{aa7})] truncates the curves in panel~(a) before the shadow can shrink to zero. At $\eta=0$ and $a=0$ the standard Schwarzschild value $R_{\rm sh}=3\sqrt{3}\,M\approx 5.196\,M$ is recovered.}
\label{fig:9}
\end{figure*}

\begin{figure*}[ht]
\centering
    \includegraphics[width=0.45\linewidth]{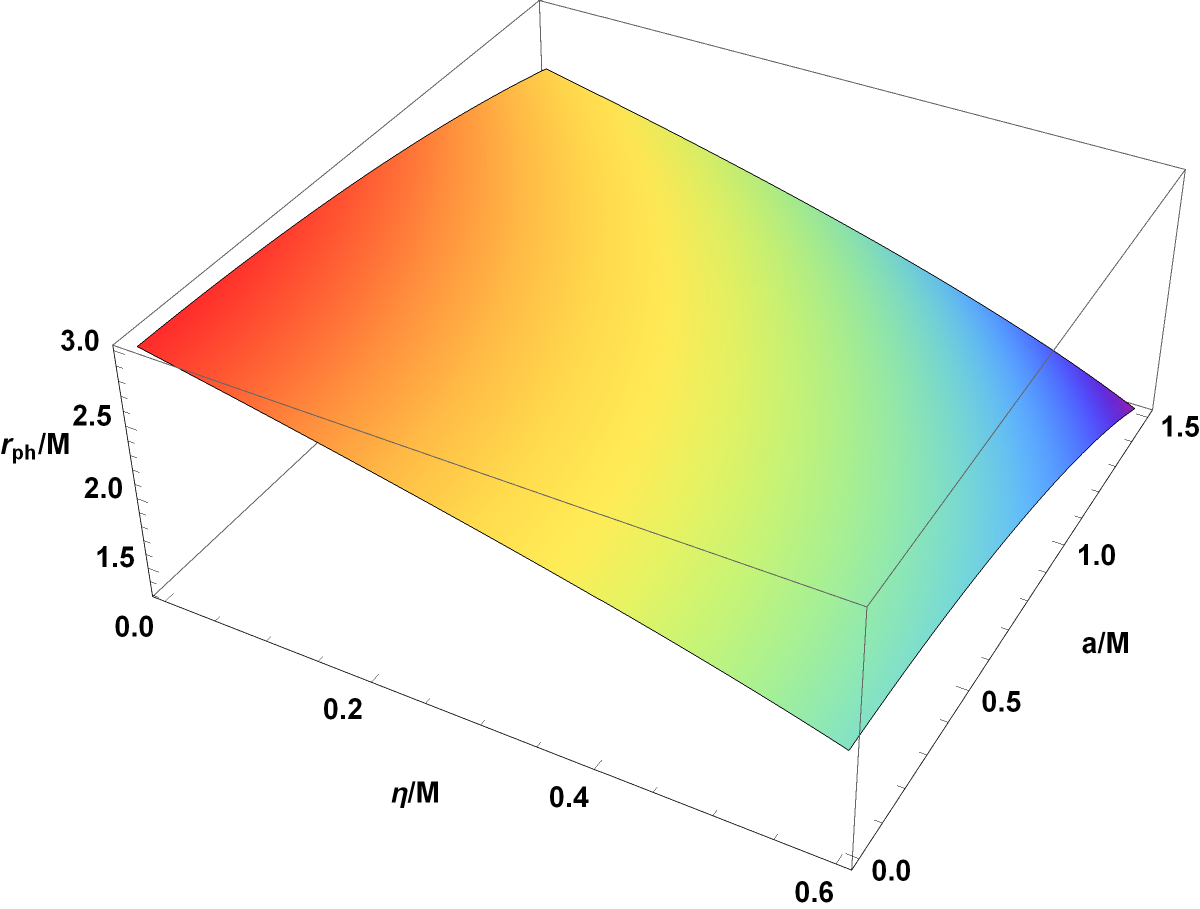}\quad\quad
\includegraphics[width=0.45\linewidth]{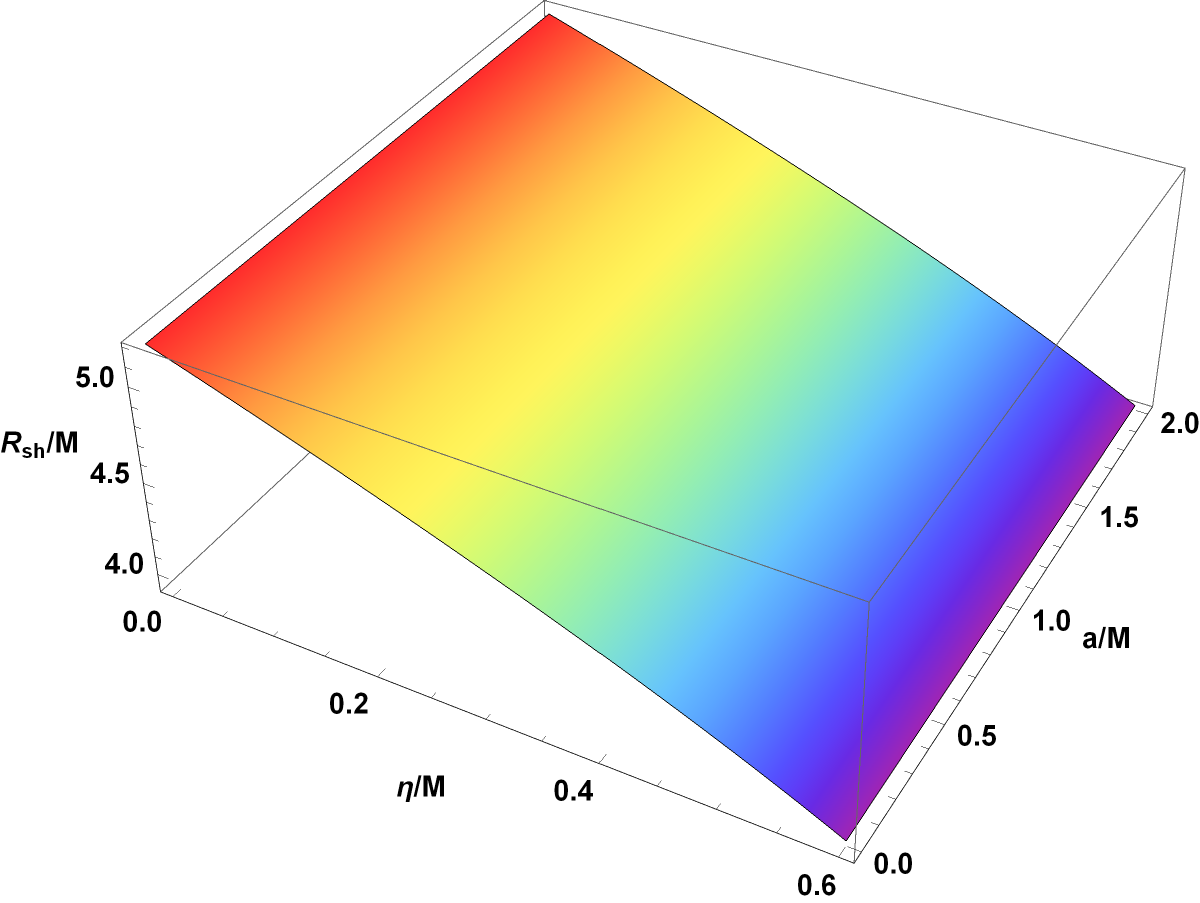}
\caption{Three-dimensional plot of the photon sphere $r_{ph}/M$ and shadow radius $R_{\rm sh}/M$ as a function of the model parameters for the SSV black hole.}
\label{fig:9a}
\end{figure*}

Circular photon orbits are determined by the condition $V'_{\text{eff}}(r) = 0$:
\begin{equation}
    r^2+a^2+\left(-3\sqrt{r^2+a^2}+\eta\right)\,M\, e^{-\eta/\sqrt{r^2+a^2}}=0.\label{ss6}
\end{equation}
In the limit $\eta=0$, one recovers the photon sphere location \(r_{\text{ph}}
=\sqrt{9M^2-a^2}.\)

Since the spacetime is asymptotically flat, the shadow radius for a distant observer is \cite{Perlick2022}:
\begin{equation}
    R_{\rm sh}=\sqrt{\frac{r_{\rm ph}^2+a^2}{1-\frac{2 M}{\sqrt{r^{2}_{\rm ph}+a^{2}}}\,\exp\left(-\frac{\eta}{\sqrt{r^2_{\rm ph}+a^2}}\right)}},\label{ss8}
\end{equation}
where $r_{\rm ph}$ is determined from Eq.~(\ref{ss6}).

In Table \ref{tab:Photon-Shadow-radius}, the numerical values of the photon sphere radius $r_{\rm ph}$ and the corresponding shadow radius $R_{\rm sh}$ for different values of the parameters $a$ and $\eta$ were presented. From this table, it can be observed that for a fixed value of $\eta$, the photon sphere radius $r_{\rm ph}$ decreases slightly as the parameter $a$ increases. However, the corresponding shadow radius $R_{\rm sh}$ remains almost unchanged, indicating that the parameter $a$ produces a weak correction to the photon sphere location and has negligible influence on the shadow size \cite{Sohan,BadawiSV}. In contrast, for a fixed value of $a$, both the photon sphere radius and the shadow radius decrease noticeably as the parameter $\eta$ increases, indicating that $\eta$ strongly affects both the photon sphere and the shadow. Furthermore, when $a$ and $\eta$ vary simultaneously, the overall variation is primarily controlled by $\eta$. Consequently, the observed decrease in both $r_{\rm ph}$ and $R_{\rm sh}$ is mainly controlled by $\eta$, while the influence of $a$ remains comparatively very weak.

Figure~\ref{fig:9} shows the shadow radius $R_{\rm sh}/M$ as a function of both model parameters in two complementary panels. Figure~\ref{fig:9}(a) displays $R_{\rm sh}/M$ as a function of the screening parameter $\eta/M$ for five fixed values of the regularization parameter $a/M$, while panel~(b) shows $R_{\rm sh}/M$ as a function of $a/M$ for five fixed values of $\eta/M$.

In Fig. \ref{fig:9}(a), the shadow radius decreases monotonically as $\eta$ increases toward its upper bound $\eta_{\rm max}=2Me^{-1}$ [Eq.~(\ref{aa7})], beyond which no real horizon exists.  Physically, the exponential screening of the gravitational potential reduces the effective photon capture cross-section, yielding a smaller apparent shadow. At $\eta=0$, the shadow radii recover the known Simpson--Visser values; in particular, the Schwarzschild limit $R_{\rm sh}=3\sqrt{3}\,M\approx 5.196\,M$ is recovered for $a=\eta=0$.  Each curve is truncated at a different value of $\eta$ because the maximum allowed screening depends on $a$ through the horizon condition: larger $a$ shifts the truncation point.

Figure \ref{fig:9}(b) shows that $R_{\rm sh}$ increases monotonically with $a$ at every fixed $\eta$, since a larger wormhole throat increases the effective areal radius of the photon orbit. The curves are uniformly shifted downward as $\eta$ increases, confirming that the two parameters act independently on the shadow: $\eta$ suppresses $R_{\rm sh}$ through exponential screening, while $a$ enhances it through the geometric enlargement of the photon sphere (see also Figure \ref{fig:9a}).  The joint sensitivity of the shadow radius to both parameters makes it an observable discriminator between the SSV geometry and general relativity, with current Event Horizon Telescope constraints providing direct bounds on the allowed region of the $(a,\eta)$ parameter space. 

The effective radial force experienced by photon particles is
\begin{equation}
    \mathrm{F}_{\rm ph}=-\frac{1}{2}\,\frac{\partial V^{\rm null}_{\rm eff}}{\partial r}.\label{ss9}
\end{equation}
Using the potential in (\ref{null-potential}), one obtains
\begin{align}
    \mathrm{F}_{\rm ph}&=-\frac{\mathrm{L}^2 r}{(r^2+a^2)^2}
\Bigg[
\frac{M}{\sqrt{r^2+a^2}}
e^{-\frac{\eta}{\sqrt{r^2+a^2}}}\left(1-\frac{\eta}{\sqrt{r^2+a^2}}\right)\nonumber\\
&-\Bigg\{
1-\frac{2M}{\sqrt{r^2+a^2}}e^{-\frac{\eta}{\sqrt{r^2+a^2}}}
\Bigg\}
\Bigg].\label{ss10}
\end{align}

\subsection{Particle Dynamics}

For massive particle dynamics, the effective potential simplifies to
\begin{equation}
    V^{\rm time-like}_{\rm eff}=\left(1+\frac{\mathcal{L}^2}{r^2+a^2}\right)\,A(r),\label{null-potential2}
\end{equation}
with the equation of motion
\begin{equation}
\left(\frac{dr}{d\lambda}\right)^2+V^{\rm time-like}_{\rm eff}=\mathcal{E}^2,\label{motion}
\end{equation}
where $\mathcal{L}$ is the conserved angular momentum per unit mass and $\mathcal{E}$ is the energy per unit mass.

\begin{figure}[ht!]
\centering
\includegraphics[width=0.95\linewidth]{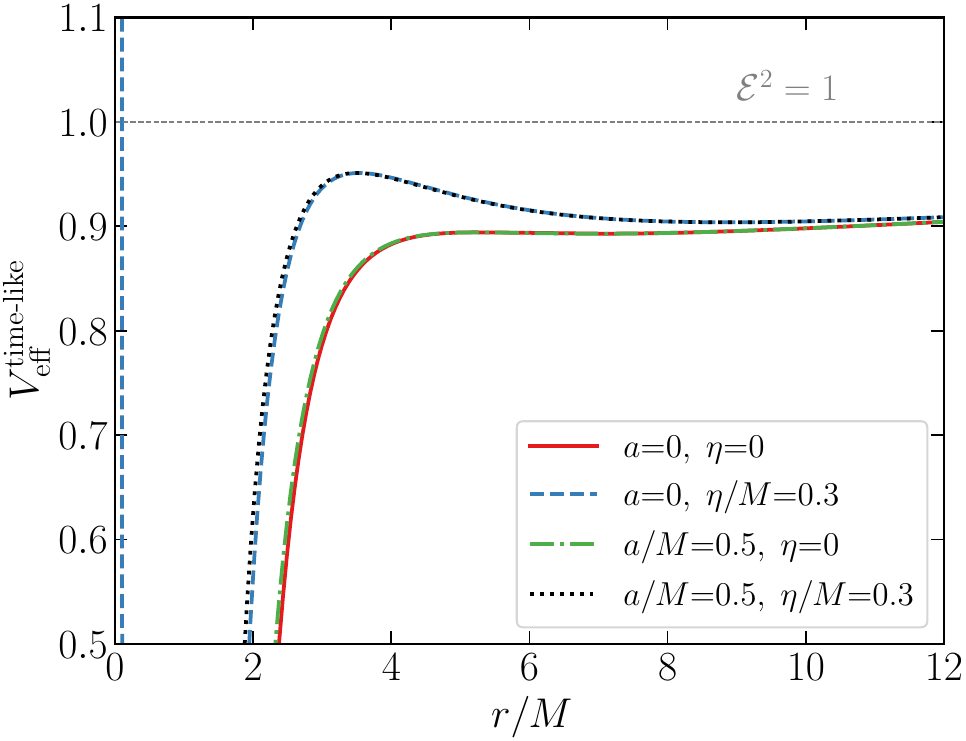}
\caption{Time-like effective potential $V^{\rm time-like}_{\rm eff}$ as a function of $r/M$ for $\mathcal{L}=3.5\,M$ and the parameter sets $(a,\,\eta)=(0,0)$ (black dotted), $(0,\,0.3M)$ (red solid), $(0.5M,\,0)$ (blue dashed), and $(0.5M,\,0.3M)$ (green dash-dot).  The local maximum of each curve marks the unstable circular orbit (light ring), while the local minimum marks the ISCO.  The horizontal dashed line at $\mathcal{E}^2=1$ indicates the marginally bound orbit.  Both deformation parameters shift the ISCO relative to the Schwarzschild value $r_{\rm ISCO}=6M$: the screening parameter $\eta$ moves it inward, while the regularization parameter $a$ moves it outward.  The combined effect (green dash-dot) reflects the competition between these two tendencies.}
\label{fig:8}
\end{figure}

Figure~\ref{fig:8} shows $V^{\rm time-like}_{\rm eff}$ as a function of $r/M$ for $\mathcal{L}=3.5M$.  Each curve displays a local maximum (the location of the unstable circular orbit) followed by a local minimum (the innermost stable circular orbit, ISCO), and then rises toward the origin.  The standard Schwarzschild potential ($a=\eta=0$, dotted) has its ISCO at
$r_{\rm ISCO}=6M$.  The screening parameter $\eta$ (red curve) shifts the local minimum inward, indicating a smaller ISCO.  The regularization parameter $a$ (blue curve) displaces the minimum
outward slightly, consistent with the enlarged effective radial scale $\rho=\sqrt{r^2+a^2}$.  The doubly-deformed case (green curve) produces the most modified potential landscape.  The dashed
horizontal line at $\mathcal{E}^2=1$ marks the marginally bound orbit; orbits below this line are gravitationally bound.

For circular orbits, the following conditions must be satisfied:
\begin{align}
    &\mathcal{E}^2=V^{\rm time-like}_{\rm eff}=\left(1+\frac{\mathcal{L}^2}{r^2+a^2}\right)\,A(r),\label{ss11}\\
    &\partial_r V^{\rm time-like}_{\rm eff}=0.\label{ss12}
\end{align}

Using potential (\ref{null-potential2}), we find the specific angular momentum as
\begin{align}
    \mathcal{L}^2_{\rm sp}= \frac{(r^2+a^2)\, A'(r)}{\frac{2 r}{r^2+a^2}\, A(r) - A'(r)},\label{ss13}
\end{align}
and the specific energy as
\begin{equation}
\mathcal{E}^2_{\rm sp} = \frac{2 r\, A(r)^2}{2 r\, A(r) - (r^2+a^2)\, A'(r)},\label{ss14}
\end{equation}
where
\begin{equation}
A'(r)= \frac{2 M r}{(r^2+a^2)^{3/2}} \exp\!\left(-\frac{\eta}{\sqrt{r^2+a^2}}\right) \left(1-\frac{\eta}{\sqrt{r^2+a^2}}\right).\label{ss15}
\end{equation}

For stable circular orbits, we additionally require:
\begin{equation}
\partial^2_r V^{\rm time-like}_{\rm eff} \geq 0.\label{ss16}
\end{equation}

For a marginally stable circular orbit, where $\partial^2_r V^{\rm time-like}_{\rm eff}=0$, we find
\begin{equation}
\frac{r\, A(r)\, A''(r)}{A'(r)} - 2 r\, A'(r) + \frac{(3 r^2 - a^2)\, A(r)}{r^2+a^2} = 0,\label{ss17}
\end{equation}
where
\begin{align}
A''(r) &= \frac{2 M \, e^{-\eta/\sqrt{r^2+a^2}}}{(r^2+a^2)^{5/2}}
\Bigg[ (a^2 - 2 r^2) \left(1 - \frac{\eta}{\sqrt{r^2+a^2}}\right) \notag\\&+ r^2 \eta \left( 2 - \frac{\eta}{\sqrt{r^2+a^2}} \right) \Bigg].\label{ss18}
\end{align}    

The exact analytical solution of Eq.~(\ref{ss17}) yields the ISCO radii.  However, due to the presence of the exponential term, a closed-form solution is highly challenging.  We therefore compute
the ISCO radii numerically. In Table~\ref{tab:ISCO-radius}, we present the numerical values of the ISCO radius for different choices of the parameters $a$ and $\eta$. From the table, it can be observed that for a fixed value of $\eta$, the position of the ISCO radius $r_{\rm ISCO}$ gradually decreases as the parameter $a$ increases. A similar trend is also observed when $a$ is held fixed and the parameter $\eta$ increases. This behavior indicates that both parameters $a$ and $\eta$ play a significant role in determining the location of the ISCO in test-particle motion around the black hole.

\begin{figure}[ht]
\centering
\includegraphics[width=0.95\linewidth]{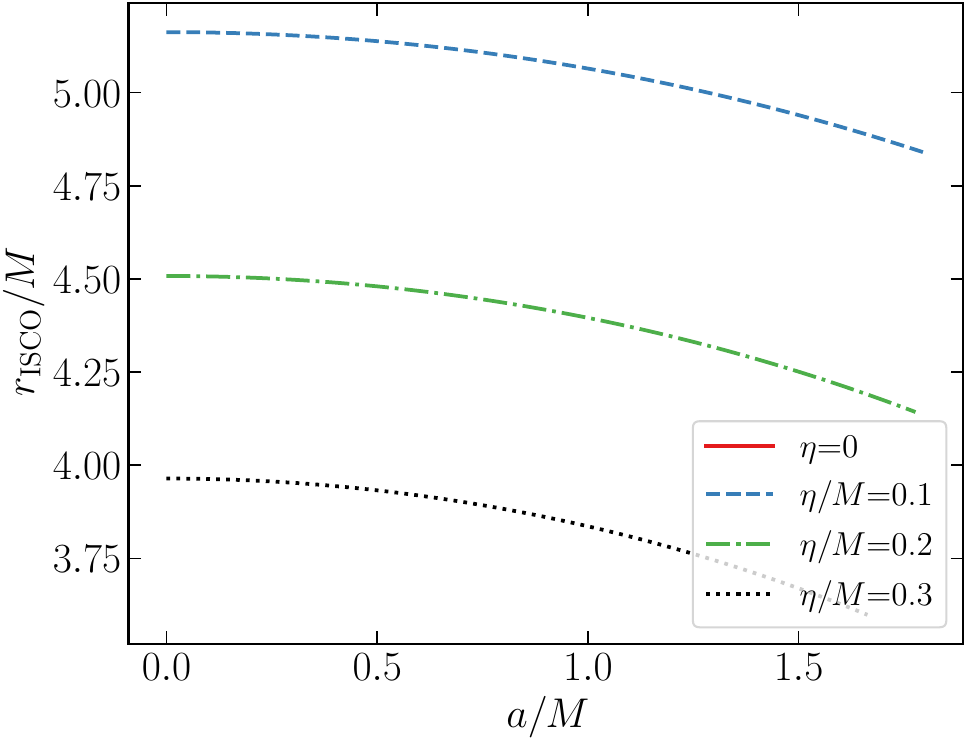}
\caption{ISCO radius $r_{\rm ISCO}/M$ as a function of the regularization parameter $a/M$ for $\eta/M=0$ (red solid), $0.1$ (blue dashed), $0.2$ (green dash-dot), and $0.3$ (black dotted).  At $a=\eta=0$, the standard Schwarzschild value $r_{\rm ISCO}=6M$ is recovered.  Increasing $a$ decreases the ISCO radius; the curves terminate when $a$ becomes large enough that the ISCO merges with the photon sphere and the geometry transitions to a wormhole.  The screening parameter $\eta$ shifts the ISCO inward at every $a$, reflecting the reduced gravitational binding induced by the exponential suppression of the gravitational potential.  This behavior directly affects the accretion disk structure and the observable radiation spectrum.}
\label{fig:11}
\end{figure}

\begin{figure}[ht]
\centering
\includegraphics[width=0.95\linewidth]{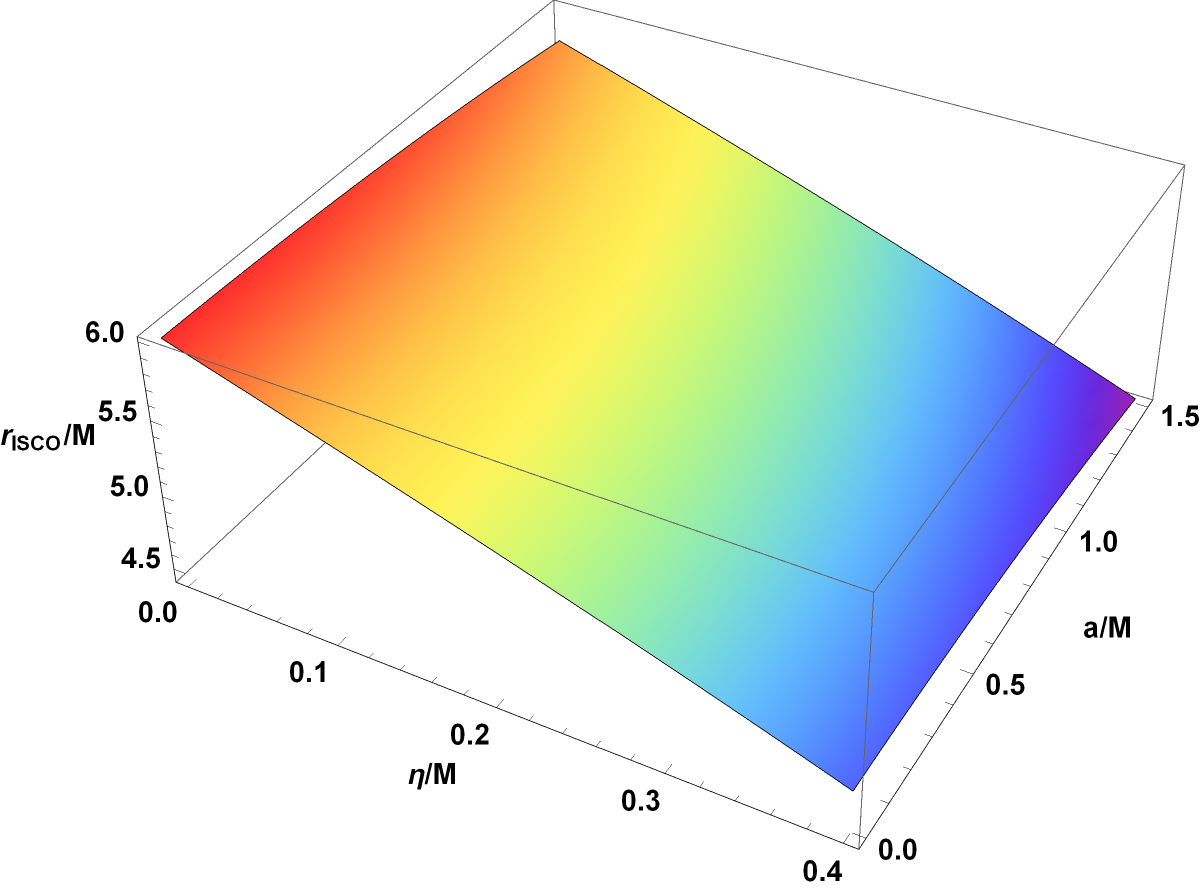}
\caption{Three dimensions plot of the ISCO radius $r_{\rm ISCO}/M$ as a function of the  parameters $a/M$ and $\eta/M$.}
\label{fig:11a}
\end{figure}

Figure~\ref{fig:11} shows the ISCO radius $r_{\rm ISCO}/M$ as a function of the regularization parameter $a/M$ for four values of $\eta$.  The Schwarzschild value
$r_{\rm ISCO}=6M$ is recovered for $a=\eta=0$.  Increasing $a$ decreases $r_{\rm ISCO}$ smoothly; at sufficiently large $a$, the ISCO merges with the photon sphere, signaling the transition from a
black hole to a traversable wormhole.  The screening parameter $\eta$ modulates this behavior: larger $\eta$ shifts the ISCO curves downward (inward) at every $a$, reflecting the weakened gravitational binding due to exponential screening (see also Figure \ref{fig:11a}).  Accurate knowledge of the ISCO radius is observationally relevant because it sets the inner edge of the accretion disk and thereby influences the emitted radiation spectrum.

\begin{table}[h!]
\centering
\begin{tabular}{|c|c|c|c|c|c|}
\hline
$a/M \backslash \eta/M$ & 0.1 & 0.2 & 0.3 & 0.4 & 0.5 \\
\hline
0.1 & 5.6935 & 5.3750 & 5.0405 & 4.68536 & 4.30214 \\
0.2 & 5.69087 & 5.3722 & 5.03752 & 4.68216 & 4.29865 \\
0.3 & 5.68647 & 5.36755 & 5.03256 & 4.67682 & 4.29283 \\
0.4 & 5.68032 & 5.36102 & 5.0256 & 4.66933 & 4.28467 \\
0.5 & 5.67239 & 5.35262 & 5.01664 & 4.65968 & 4.27416 \\
\hline
\end{tabular}
\caption{Numerical values of $r_{\rm ISCO}/M$ for different values of $a$ and $\eta$ with $M=1$.}
\label{tab:ISCO-radius}
\end{table}


Finally, the effective radial force experienced by massive test particles is
\begin{align}
\mathcal{F}&=-\frac{A'(r)}{2}-\frac{\mathcal{L}^2 r}{(r^2+a^2)^2}
\Bigg[
\frac{M}{\sqrt{r^2+a^2}}
\exp\!\left(-\frac{\eta}{\sqrt{r^2+a^2}}\right)\notag\\& \times
\left(1-\frac{\eta}{\sqrt{r^2+a^2}}\right)
-
\Bigg\{
1-\frac{2M}{\sqrt{r^2+a^2}}
\notag\\ &\times\exp\!\left(-\frac{\eta}{\sqrt{r^2+a^2}}\right)
\Bigg\}
\Bigg].\label{ss19}
\end{align}

\section{Energy Emission Rate}\label{sec:5}

In the high-frequency (geometric-optics) limit, the absorption cross section exhibits oscillations around a constant value, denoted $\sigma_{\rm lim}$.  Since the capture of energetic quanta is largely determined by null geodesics, the black hole shadow directly influences the high-energy absorption cross section.  The constant limiting value of the cross section is \cite{Misner1973,Mashhoon1973,Wei2013}:
\begin{equation}
\sigma_{\rm lim}\approx \pi R^2_{\rm sh},\label{ee1}
\end{equation}
where \(R_{\rm sh}\) denotes the shadow radius.
Within this approximation, the spectral energy emission rate of the black hole is \cite{Wei2013}
\begin{equation}
\frac{d^{2}\mathbb{E}}{d\omega\,dt}=\frac{2\pi ^{2}\sigma _{\rm lim}}{e^{\omega/T}-1}\,\omega ^{3},\label{ee2}
\end{equation}
where $\omega$ is the emitted frequency and $T$ is the Hawking temperature.

Substituting $T$ from Eq.~(\ref{aa12}), we find
\begin{equation}
\frac{d^{2}\mathbb{E}}{d\omega\,dt}=\frac{2\pi ^{3} R^2_{\rm sh}\,\omega ^{3}}{\left[\exp\left\{\frac{4\pi \omega}{r_h}\left(\frac{1}{r_h^2+a^2}-\frac{\eta}{(r_h^2+a^2)^{3/2}}\right)^{-1}\right\}-1\right]}.\label{ee3}
\end{equation}

\begin{figure}[ht!]
\centering
\includegraphics[width=0.9\linewidth]{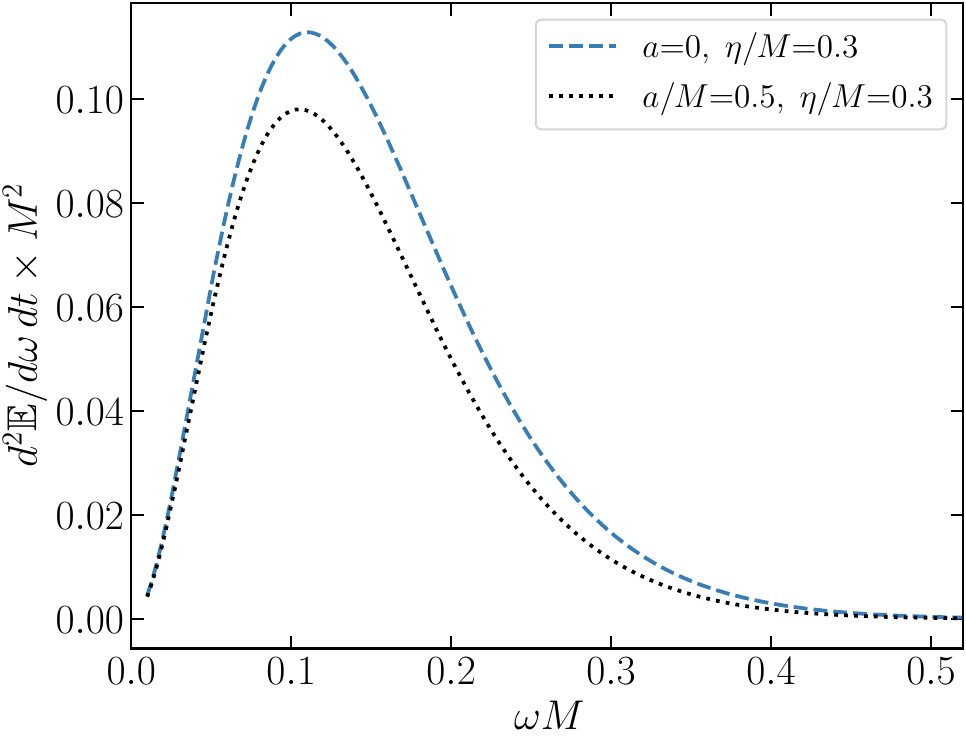}
\caption{Spectral energy emission rate $d^2\mathbb{E}/d\omega\,dt \times M^2$ as a function of $\omega M$ for $(a,\,\eta)=(0,0)$ (black dotted), $(0,\,0.3M)$ (red solid),
$(0.5M,\,0)$ (blue dashed), and $(0.5M,\,0.3M)$ (green dash-dot).  Each curve exhibits the characteristic Planckian peak, whose height and position encode the geometry's Hawking temperature and shadow radius.  Both $\eta$ and $a$ lower $T$ and $R_{\rm sh}$, reducing the emitted power and shifting the spectral peak toward lower frequencies. The doubly-deformed case (green dash-dot) produces the most strongly suppressed and red-shifted emission.  These modifications to the Hawking spectrum may serve as observational signatures of the screened SV geometry.}
\label{fig:10}
\end{figure}

Figure~\ref{fig:10} displays the spectral energy emission rate $d^2\mathbb{E}/d\omega\,dt$ as a
function of the dimensionless frequency $\omega M$ for four parameter combinations.  Each curve exhibits the characteristic Planckian shape: a rise at low frequencies, a peak at an intermediate
frequency $\omega_{\rm peak}$, and a rapid fall-off at high frequencies.  For the Schwarzschild case ($a=\eta=0$, dotted), the peak is highest and located at the largest frequency, reflecting the highest Hawking temperature.  Introducing the screening parameter $\eta$ (red curve) lowers $T$, which shifts the peak to lower frequencies and reduces the total power radiated, in accordance with
Wien's displacement law.  The regularization parameter $a$ produces a similar shift but with a different functional dependence: the peak moves further downward in both amplitude and frequency. The doubly-deformed case ($a/M=0.5,\,\eta/M=0.3$) yields the lowest and most red-shifted peak among the four cases, signaling the most strongly suppressed emission.  These differences in the spectral profile are potentially observable through future near-horizon measurements or analog gravity experiments.

\section{Photon Sphere Topology}\label{sec:6}

In order to study the topological property of the photon sphere (PS), one can first introduce a potential function \cite{PVPC1,PVPC2,PVPC3,PVPC4}:
\begin{equation}
    H(r,\theta)=\sqrt{-\frac{g_{tt}}{g_{\phi\phi}}}=\frac{1}{\sin \theta}\sqrt{\frac{A(r)}{r^2+a^2}},\label{qq1}
\end{equation}
where $H(r,\theta)$ is regular. The photon sphere radius is determined by the root of $\partial_r H(r,\theta)=0$. Following Ref.~\cite{PVPC1}, one considers the general vector field ${\bf v}=(v_r,v_\theta)$ defined by
\begin{equation}
v_r=\frac{\partial_r H(r,\theta)}{\sqrt{g_{rr}}}\quad,\quad v_\theta=\frac{\partial_\theta H(r,\theta)}{\sqrt{g_{\theta\theta}}}.\label{qq2}
\end{equation}
For the metric (\ref{metric}),
\begin{equation}
g_{rr}=\frac{1}{A(r)}\quad,\quad g_{\theta\theta}=(r^2+a^2)\,\sin^2 \theta.\label{qq3}
\end{equation}
Substituting $H(r,\theta)$ from Eq.~(\ref{qq1}) and using Eq.~(\ref{qq3}) into Eq.~(\ref{qq2}) gives
\begin{align}
 v_r&=\frac{A'(r)(r^2+a^2)-2rA(r)}
{2\sin\theta\,(r^2+a^2)^{3/2}},\label{qq4}\\
v_\theta&=-\frac{\cos\theta}{\sin^3\theta}\,
\frac{\sqrt{A(r)}}{r^2+a^2}.\label{qq5}
\end{align}
The vector ${\bf v}$ can be written as
\begin{equation}
{\bf v}=v\,e^{i\,\Omega},\quad v=|{\bf v}|.\label{qq7}
\end{equation}
The zero point of ${\bf v}$ coincides exactly with the photon sphere location $(r,\theta)=(r_{\rm ph},\pi/2)$, so ${\bf v}$ is not well-defined there.  The normalized vector field is
\begin{equation}
 {\bf n}=(n_r,n_{\theta})=\frac{{\bf v}}{v},\label{qq8}
\end{equation}
with
\begin{align}
n_r=\frac{v_r}{\sqrt{v^2_r+v^2_{\theta}}}\quad,\quad n_{\theta}=\frac{v_{\theta}}{\sqrt{v^2_r+v^2_{\theta}}}.\label{qq9}
\end{align}

The topological current is formulated using Duan's $\phi$-mapping topological current theory \cite{duan1998,duan1999}:
\begin{equation}
j^{\mu}=\frac{1}{2\,\pi}\,\varepsilon^{\mu\nu\rho}\,\varepsilon_{ab}\,\partial_{\nu}\,n^a\,\partial_{\rho}\,n^b,\quad \mu,\nu,\rho=0,1,2,\label{qq10}
\end{equation}
with $\partial_{\mu} j^{\mu}=0$.  The topological charge over region $\Sigma$ is
\begin{equation}
Q=\int_{\Sigma} j^0\, d^2x = \sum_{i=1}^{N} \beta_i \eta_i = \sum_{i=1}^{N} w_i.\label{qq11}
\end{equation}
Here $\beta_i$ is the positive Hopf index, $\eta_i=\mathrm{sign}(J^0(n/x)_{z_i})=\pm 1$ is the Brouwer degree, and the winding number around the $i$-th zero is
\begin{equation}
w_i = \frac{1}{2\pi} \oint_{C_i} d\Omega,\label{qq12}
\end{equation}
where $\Omega = \arctan\!\left(\frac{v_\theta}{v_r}\right)$.

\begin{figure}[ht!]
\centering
\includegraphics[width=0.9\linewidth]{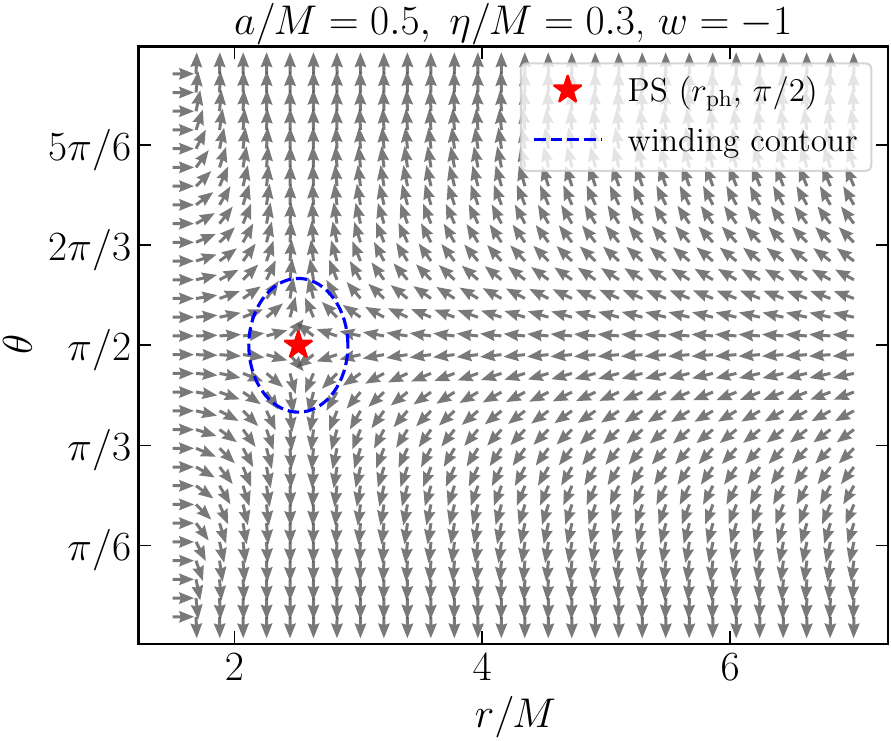}
\caption{Topological structure of the photon sphere for $a/M=0.5$, $\eta/M=0.3$, displayed as the normalized vector field $\mathbf{n}=(n_r,n_\theta)$ in the $(r/M,\,\theta)$ plane.  The red star marks the zero of $\mathbf{n}$ at the photon sphere location $(r_{\rm ph},\pi/2)$.  The blue dashed circle is the winding-number contour $C$ enclosing this zero; integrating $d\Omega$ around $C$ yields a winding number $w=-1$, confirming the photon sphere as a topological defect with negative unit charge.  The arrows show the direction of $\mathbf{n}$; their reversal across $\theta=\pi/2$ reflects the equatorial mirror symmetry.  The topological charge $w=-1$ is a universal invariant that persists for any smooth choice of $a$ and $\eta$ within the black hole parameter range.}
\label{fig:12}
\end{figure}

Figure~\ref{fig:12} illustrates the topological structure of the photon sphere in the $(r/M, \theta)$ plane for the representative parameter set $a/M=0.5$, $\eta/M=0.3$. The arrows depict the
normalized vector field $\mathbf{n}=(n_r, n_\theta)$ evaluated on a regular grid covering the exterior of the black hole. The red star marks the zero point of $\mathbf{n}$, which coincides exactly with the photon sphere location $(r_{\rm ph},\pi/2)$, the equatorial circular null geodesic. As the field $\mathbf{n}$ is traced around the small closed contour (blue dashed circle) enclosing this zero point, the vector direction rotates by a net angle of $-2\pi$, corresponding to a winding number $w=-1$.  This topological charge is a robust, parameter-independent invariant: it is preserved under smooth deformations of $a$ and $\eta$ as long as no additional zero points are created or annihilated.  The $\theta$-component $n_\theta$ changes sign across the equatorial plane $\theta=\pi/2$, enforcing the equatorial mirror symmetry of the vector field and confirming that the photon sphere is an isolated topological defect of the normalized field.  The stability classification based on the sign of the winding number ($w=-1$ for an unstable photon sphere) is in full agreement with the standard stability analysis of the effective potential maximum shown in Fig.~\ref{fig:7}.

\section{Thermodynamic Topology}\label{sec:7}

In Section \ref{sec:3}, we analyzed the thermodynamic properties of the considered black hole, including the temperature, Gibbs free energy, and specific heat. The phase structure can be further examined from a topological perspective using the method of Wei \textit{et al.}~\cite{PVPC4,PVPC5}, which treats black hole configurations as topological defects in an auxiliary parameter space and characterizes their stability via winding numbers. This approach has been successfully applied to various black holes in general relativity and modified gravity. Here, we employ the generalized Gibbs free energy formalism of Wei \textit{et al.}~\cite{PVPC4,PVPC5} to study the influence of geometric parameters on the thermodynamic topology.

\begin{figure*}[!t]
\centering
\includegraphics[width=0.9\linewidth]{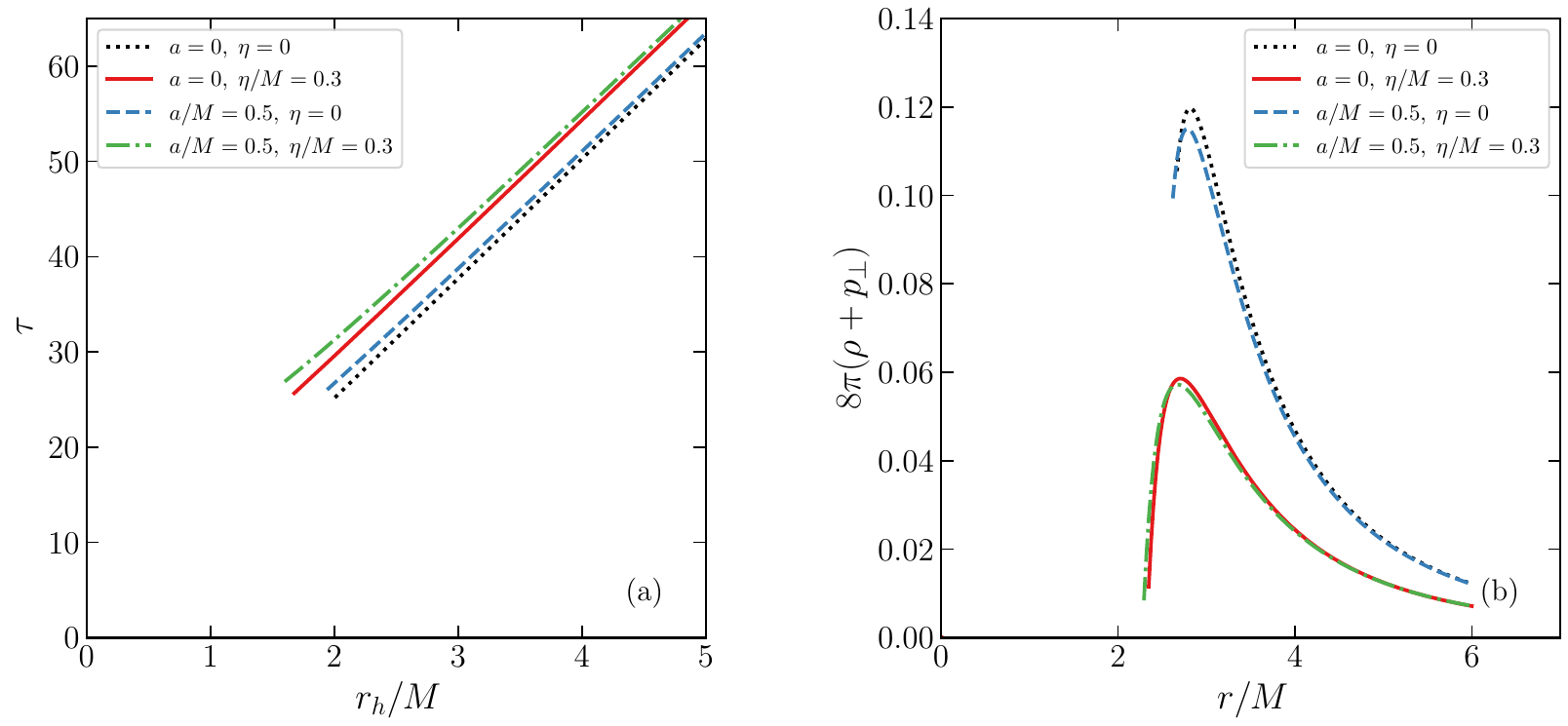}
\caption{Panel~(a): Euclidean time $\tau$ from Eq.~(\ref{time}) as a function of $r_h/M$ for $(a,\,\eta)=(0,0)$ (black dotted),
$(0,\,0.3M)$ (red solid), $(0.5M,\,0)$ (blue dashed), and
$(0.5M,\,0.3M)$ (green dash-dot). In the Schwarzschild limit $\tau=4\pi r_h=1/T_{\rm Sch}$ is recovered (black dotted).  Both $a$ and $\eta$ suppress $\tau$ and displace its zero point, the locus of the thermodynamic phase transition, to larger $r_h$, in agreement with the Davies-type divergence of the specific heat shown in Fig.~\ref{fig:5}. Panel~(b): Energy condition combination $8\pi(\rho+p_{\perp})$ from Eq.~(\ref{condition}) as a function of $r/M$ for the same four parameter sets.  The horizontal gray line marks the threshold for violation of the null energy condition (NEC) in the tangential sector. The standard Schwarzschild case ($a=\eta=0$) satisfies $8\pi(\rho+p_{\perp})>0$ throughout the exterior, while the introduction of $a$ or $\eta$ drives the combination negative at small $r$, signaling NEC violation in that regime, a generic feature of regular black hole spacetimes.  At large $r$ all curves approach zero, consistent with the asymptotic flatness of the SSV geometry. Note that $8\pi(\rho+p_r)=0$ identically [Eq.~(\ref{condition})], and $8\pi(\rho+p_r+2p_{\perp})=16\pi p_{\perp}=2\times 8\pi(\rho+p_{\perp})$, so panel~(b) encodes the complete information on the non-trivial energy condition components.}
\label{fig:tau_energy}
\end{figure*}

Given the relationship between mass and energy in black holes, the generalized free energy can be expressed as a standard thermodynamic function in the form~\cite{PVPC4,PVPC5}
\begin{equation}
\mathcal{F}=M(r_{h})-\frac{S}{\tau}, \label{ff1}
\end{equation}
where $\tau$ denotes the Euclidean time (interpreted as the inverse temperature at the zero points) outside the horizon, and $S$ is the entropy of the black hole.

In our case at hand, we find 
\begin{align}
\mathcal{F}&=\frac{\sqrt{r^2+a^2}}{2}\exp\!\left(\frac{\eta}{\sqrt{r^2+a^2}}\right)\nonumber\\
&-\frac{\pi}{\tau}\Bigg[r_h\sqrt{r_h^2+a^2}+
(a^2+\eta^2 )\ln\!\left(r_h+\sqrt{r_h^2+a^2}\right)\notag\\&+\eta \left( 2\, r_h +\frac{\eta^2}{3a}\arctan\!\left(\frac{r_h}{a}\right)\right)\Bigg].\label{ff2}
\end{align}
The components of the vector field $\boldsymbol{\phi}_{\mathcal{F}}$ associated with the generalized free energy are defined as~\cite{PVPC4,PVPC5}
\begin{align}
\phi_{\mathcal{F}}^{r_{h}}&= \partial_{r_{h}}\mathcal{F}, \nonumber\\
 \phi_{\mathcal{F}}^{\theta}&= -\cot\theta\,\csc\theta. \label{ff3}
\end{align}
Its magnitude is normalized through
\begin{equation}
\|\phi_{\mathcal{F}}\| = \sqrt{ \left( \phi_{\mathcal{F}}^{r_{h}} \right)^{2} + \left( \phi_{\mathcal{F}}^{\theta} \right)^{2} }, \label{ff4}
\end{equation}
which allows us to introduce the normalized vector field ${\bf n}_{\mathcal{F}}$ as
\begin{align}
 n_{\mathcal{F}}^{r_{h}} = \frac{\phi^{\mathcal{F}}_{r_{h}}}{\|\phi_{\mathcal{F}}\|}, \qquad 
 n_{\mathcal{F}}^{\theta} = \frac{\phi^{\mathcal{F}}_{\theta}}{\|\phi_{\mathcal{F}}\|}. \label{ff5}
\end{align} 

Now, we determine Euclidean time $\tau$ at the zero points of the vector field $\phi_{\mathcal{F}}$. This can be determined using the conditions $\theta=\pi/2$ and $\partial_{r_{h}}\mathcal{F}=0$. This yields
\begin{align}
\tau = \frac{2 \pi \left[ \dfrac{2 r_h^2 + 2 a^2 + \eta^2}{\sqrt{r_h^2 + a^2}} + \eta \left( 2 + \dfrac{\eta^2}{3 (r_h^2 + a^2)} \right) \right]}{\dfrac{r_h}{\sqrt{r_h^2 + a^2}} \left( 1 - \dfrac{\eta}{\sqrt{r_h^2 + a^2}} \right) \exp\Big(\dfrac{\eta}{\sqrt{r_h^2 + a^2}}\Big)}.\label{time}
\end{align}

\begin{itemize}
\item When $\eta=0$, the Euclidean time simplifies as
\begin{equation}
\tau \big|_{\eta = 0} = \frac{4 \pi (r_h^2 + a^2)}{r_h}=\frac{1}{T_{\rm SV}}
\end{equation}
which is similar to the SV-geometry.
    
\item When $a=0=\eta$, the Euclidean time simplifies as
\begin{equation}
\tau \big|_{a= 0,\,\eta=0} =4\pi r_h=\frac{1}{T_{\rm Sch}},
\end{equation} 
where $r_h=2M$. This is exactly what we expect for a Schwarzschild black hole, consistent with the standard Hawking temperature relation
\end{itemize}

Figure~\ref{fig:tau_energy}(a) displays the Euclidean time $\tau$ given by Eq.~(\ref{time}) as a function of the horizon radius $r_h/M$ for the same four representative parameter sets used throughout this work. In the Schwarzschild limit ($a=\eta=0$, black dotted), $\tau$ grows linearly with $r_h$, recovering the familiar relation $\tau=4\pi r_h = 1/T_{\rm Sch}$, and its zero point at $r_h=2M$ signals the standard Hawking temperature. Introducing the screening parameter $\eta>0$ (with $a=0$, red solid) lowers
$\tau$ at every horizon size and displaces the zero point to a larger $r_h$, consistent with the reduction of the Hawking temperature shown in Fig.~\ref{fig:3}. The regularization parameter $a>0$ (with $\eta=0$, blue dashed) similarly suppresses $\tau$ relative to the Schwarzschild baseline and shifts the zero of $\tau$ outward, reflecting the enlarged effective radial scale $\rho_h=\sqrt{r_h^2+a^2}$. In the doubly-deformed case ($a/M=0.5$, $\eta/M=0.3$, green dash-dot), both effects combine: $\tau$ is most strongly suppressed, and its zero point is displaced to the largest critical radius among the four cases.  The zero of $\tau$ at a finite $r_h$ marks the thermodynamic phase transition of the black hole, in agreement with the Davies-type divergence observed in the specific heat (Fig.~\ref{fig:5}), and confirms the consistency of the thermodynamic topology analysis with the conventional thermodynamic picture.

Figure~\ref{fig:tau_energy}(b) shows the combination $8\pi(\rho+p_{\perp})$ as a function of $r/M$ for the four representative parameter sets. Since $8\pi(\rho+p_r)=0$ identically [Eq.~(\ref{condition})], the only non-trivial NEC combination in the tangential sector is $8\pi(\rho+p_{\perp})$, and the SEC combination $8\pi(\rho+p_r+2p_{\perp})=16\pi p_{\perp}$ carries the same sign information. For the Schwarzschild case ($a=\eta=0$, black dotted), the combination is positive throughout the exterior, so no energy condition is violated there. Activating either $a$ or $\eta$ drives $8\pi(\rho+p_{\perp})$ negative in an inner region near the horizon, signaling violation of the NEC in the tangential sector.  This behavior is a generic feature of regular black hole solutions: the matter content responsible for resolving the singularity must
necessarily violate at least one classical energy condition, as required by the singularity theorems of Penrose~\cite{Penrose1965} and Hawking~\cite{Hawking1970}.  At large $r$ all four curves converge to zero, consistent with the asymptotic flatness of the SSV geometry. The doubly-deformed case ($a/M=0.5$, $\eta/M=0.3$, green dash-dot) exhibits the deepest violation at small $r$, reflecting the combined effect of both deformation mechanisms on the sourcing stress-energy tensor.

\section{Sparsity}
In this section, we investigate the sparsity of Hawking radiation associated with screened
Simpson–Visser black hole solution. Although Hawking radiation possesses a thermal spectrum characterized by the Hawking temperature determined from the surface gravity at the event horizon, the emission process is not continuous in time. Instead, it occurs as a sequence of well-separated quanta, implying that the radiation flux is intrinsically sparse. The degree of sparsity can be quantified by comparing the typical thermal wavelength of the emitted quanta, $\lambda_t = 2\pi/T$, with the effective emission area $\mathcal{A}_{\rm eff}$ of the black hole. Following the formulation introduced in \cite{Hawking1975,Page1976,Gray2016}, the sparsity of the Hawking radiation is characterized by a dimensionless parameter defined as
\begin{equation}
    \tilde{\eta} =\frac{\mathcal{C}}{\Tilde{g} }\left(\frac{\lambda_t^2}{\mathcal{A}_{\rm eff}}\right),\label{Sparsity-1}
\end{equation}
where \(\mathcal{C}\) is a dimensionless constant, \(\tilde g\) the spin degeneracy of the emitted quanta and \(\mathcal{A}_{\rm eff}=\frac{27}{4}\,A_{\rm BH}\) is the effective area.

For the considered black hole, using the temperature $T$ in (\ref{aa12}) and the BH area in (\ref{area}), we find the dimensionless sparsity parameter
\begin{equation}
    \tilde{\eta}=\left(1+\frac{a^2}{r_h^2}\right)\left(1-\frac{\eta}{\sqrt{r_h^2+a^2}}\right)^{-2}\,\tilde{\eta}_{\rm Sch.},\label{Sparsity-3}
\end{equation}
where $\tilde{\eta}_{\rm Sch.}=64\pi^3/27 \approx 73.5$.

\begin{itemize}
    \item When $\eta=0$, the sparsity parameter simplifies as
    \begin{equation}
        \tilde{\eta}=\left(1+\frac{a^2}{r_h^2}\right)\,\tilde{\eta}_{\rm Sch},\quad r_h=\sqrt{4 M^2-a^2}.\label{Sparsity-4}
    \end{equation}

    \item When $a=0$, the sparsity parameter simplifies as
    \begin{equation}
    \tilde{\eta}=\left(1-\frac{\eta}{r_h}\right)^{-2}\,\tilde{\eta}_{\rm Sch.},\quad r_h=\rho_h,\label{Sparsity-5}
\end{equation}
where $\rho_h$ is given in (\ref{aa5}).
\end{itemize}

Our analysis shows that the sparsity of Hawking radiation for the screened SV black hole is highly sensitive to the regularization parameters ($a, \eta$), leading to deviations from the behavior observed in the standard Schwarzschild black hole.

\section{Conclusions}\label{sec:8}

In this paper, we have introduced the screened Simpson--Visser (SSV) black hole, a static, spherically symmetric regular spacetime constructed by merging the exponential screening prescription of Ref.~\cite{Alex2020} with the wormhole-regularization of Ref.~\cite{Simpson2019}.  The resulting two-parameter metric function simultaneously resolves the central curvature singularity and modifies the gravitational potential at all radial scales,
interpolating continuously between the two parent models and recovering standard Schwarzschild geometry in the limit $a, \eta \to 0$.

The computation of the Ricci scalar, the Kretschmann invariant, and the squared Ricci tensor showed that all curvature scalars are finite throughout the spacetime, including at the origin, establishing the global regularity of the SSV geometry.  The horizon structure, determined analytically through the Lambert $W$ function, yields an outer event horizon and an inner Cauchy horizon whenever $\eta < 2Me^{-1}$; outside this bound the geometry transitions to a one-way and, for sufficiently large $a$, to a two-way traversable wormhole.  These features demonstrate that the two deformation parameters together provide precise control over the causal character of the spacetime.

The thermodynamic analysis revealed that both $a$ and $\eta$ reduce the Hawking temperature compared to the Schwarzschild case, with the screening parameter further enforcing a zero-temperature extremal limit.  The entropy satisfies the Bekenstein--Hawking area law, growing quadratically with the horizon radius for any fixed $a$, while $\eta$ affects the entropy only through its displacement of the horizon.  A Davies-type phase transition is present in the specific heat at a critical horizon radius that shifts outward with increasing $a$ or $\eta$, separating a small, thermally unstable black hole phase from a large, thermally stable one.  Throughout the accessible temperature range, the Helmholtz free energy remains positive, indicating the absence of a Hawking--Page transition and suggesting enhanced thermodynamic stability compared to the Schwarzschild geometry.

The geodesic structure reflects the same competition between the two parameters.  The photon sphere is displaced inward by $\eta$ and outward by $a$, and the shadow radius follows the same opposing
trends, recovering the Schwarzschild value $R_{\rm sh} = 3\sqrt{3}\,M$ only when both parameters vanish.  The constraint $\eta \leq 2Me^{-1}$ places a natural lower bound on the shadow size within the SSV family, making the model falsifiable by very-long-baseline interferometry observations of M87* and Sgr~A*.  For massive test particles, the ISCO exhibits the same opposing dependence on $a$ and $\eta$, with direct consequences for the accretion disk structure and the radiative efficiency of infalling matter.  The spectral energy emission rate inherits all these modifications: the Planckian peak is shifted to lower frequencies and reduced in amplitude relative to the Schwarzschild case, with the doubly-deformed case producing the most strongly suppressed spectrum and offering a potential observational discriminator between the SSV geometry and its limiting cases.

Finally, the topological analysis via Duan's $\phi$-mapping framework assigned a winding number $w = -1$ to the photon sphere for every parameter combination within the black hole regime.  This
value is a topological invariant, preserved under any smooth continuous deformation of $a$ and $\eta$, and its negative sign is consistent with the unstable nature of the photon orbit as a
maximum of the effective potential.  The universality of this result establishes the photon sphere topology as an intrinsic, parameter-independent characteristic of the SSV class of regular black holes.

Looking ahead, the most immediate extension of this work is the construction of a rotating SSV black hole, which would add frame-dragging, an ergosphere, and superradiant instabilities to the
picture.  Further natural directions include the study of weak and strong gravitational lensing, the computation of quasinormal modes and greybody factors, a systematic assessment of the energy conditions for the sourcing matter fields, and the derivation of the SSV metric from first principles within a non-commutative geometry or loop quantum gravity framework.  We leave these investigations for future work.\\

\section*{acknowledgments}
F.A. gratefully acknowledges the Inter University Center for Astronomy and Astrophysics (IUCAA), Pune, India, for the conferment of a visiting associateship. 

\section*{Data Availability Statement}
There are no new data associated with this article [Authors comment: No data were generated in this article].

\section*{Code/Software}

No code/software were developed in this article [Authors comment: No code/software were developed in this article].

\end{document}